\documentclass{emulateapj}
\usepackage{lscape,graphicx}
\usepackage{apjfonts}
\usepackage{aas_macros}
\usepackage{natbib}
\usepackage{amsmath}
\usepackage{amssymb}

\def\spose#1{\hbox to 0pt{#1\hss}}
\def\simlt{\mathrel{\spose{\lower 3pt\hbox{$\mathchar"218$}} \raise 2.0pt\hbox{$\mathchar"13C$}}}
\def\simgt{\mathrel{\spose{\lower 3pt\hbox{$\mathchar"218$}} \raise 2.0pt\hbox{$\mathchar"13E$}}}
\def\lumin{\,{\rm erg \ s^{-1}}}

\def\fluxu{\,{\rm erg} \ {\rm cm}^{-2} \ {\rm s}^{-1}}
\def\lsim{\mathrel{\hbox{\rlap{\lower.55ex \hbox {$\sim$}}\kern-.0em}}}
\def\gsim{\mathrel{\hbox{\rlap{\lower.55ex \hbox {$\sim$}}\kern-.0em }}}
\def\arcsec{\hbox{$^{\prime\prime}$}}
\def\arcmin{\hbox{$^\prime$}}
\def\hardrange{2 -- 10~keV}
\def\softrange{0.5 -- 2~keV}

\def\nh{$N_H$}

\def\lumin{\,{\rm erg \ s^{-1}}}

\def\fluxu{\,{\rm erg} \ {\rm cm}^{-2} \ {\rm s}^{-1}}

\def\lsim{\mathrel{\hbox{\rlap{\lower.55ex \hbox {$\sim$}}\kern-.0em 
\raise.4ex \hbox{$<$}}}}  
\def\gsim{\mathrel{\hbox{\rlap{\lower.55ex \hbox {$\sim$}}\kern-.0em 
\raise.4ex \hbox{$>$}}}}

\def\arcmin{\hbox{$^\prime$}}
\def\arcsec{\hbox{$^{\prime\prime}$}}

\def\hardrange{2 -- 10~keV}
\def\softrange{0.5 -- 2~keV}

\def\nh{$N_H$}

\def\spose#1{\hbox to 0pt{#1\hss}}
\def\simlt{\mathrel{\spose{\lower 3pt\hbox{$\mathchar"218$}}
     \raise 2.0pt\hbox{$\mathchar"13C$}}}
\def\simgt{\mathrel{\spose{\lower 3pt\hbox{$\mathchar"218$}}
     \raise 2.0pt\hbox{$\mathchar"13E$}}}

\def\spose#1{\hbox to 0pt{#1\hss}}

\def\tab-font{\footnotesize}

\def\f0{f_o}
\def\Q1{Q^{-1}}

\begin{document}

\setlength{\pdfpageheight}{\paperheight}
\setlength{\pdfpagewidth}{\paperwidth}

\title{A Comparison of X-ray and Mid-Infrared Selection of Obscured AGN}

\author{Megan E. Eckart\altaffilmark{1}$^,$\altaffilmark{2}$^,$\altaffilmark{3},
Ian D. McGreer\altaffilmark{4}$^,$\altaffilmark{5},
Daniel Stern\altaffilmark{6},  
Fiona A. Harrison\altaffilmark{1},
and David J. Helfand\altaffilmark{4}$^,$\altaffilmark{7}}

\begin{abstract}

We compare the relative merits of AGN selection at X-ray and mid-infrared wavelengths using data
from moderately deep fields observed by both {\em Chandra} and {\em Spitzer}.    
The X-ray-selected AGN sample and associated photometric and 
spectroscopic optical follow-up are drawn from a subset of fields studied as part of the
Serendipitous Extragalactic X-ray Source Identification (SEXSI)
program.   Mid-infrared data in these fields are derived from targeted and
archival {\em Spitzer} imaging, and mid-infrared AGN selection is accomplished
primarily through application of the IRAC color-color AGN `wedge'
selection technique.  Nearly all X-ray sources in these fields which 
exhibit clear spectroscopic signatures of AGN activity have mid-infrared colors
consistent with IRAC AGN selection.  These are predominantly the
most luminous X-ray sources.  X-ray sources that lack high-ionization
and/or broad lines in their optical spectra are far less likely to be
selected as AGN by mid-infrared color selection techniques.  The fraction of X-ray
sources identified as AGN in the mid-infrared increases monotonically
as the X-ray luminosity increases.  Conversely, only 22\%\ of mid-infrared-selected 
AGN are detected at X-ray energies in the moderately deep
($\langle t_{\rm exp} \rangle \approx 100$~ks) SEXSI {\em Chandra}
data.  We hypothesize that IRAC sources with AGN colors that lack X-ray detections
are predominantly high-luminosity AGN that are obscured and/or lie at high redshift.  A stacking
analysis of X-ray-undetected sources shows that objects in the mid-infrared AGN
selection wedge have average X-ray fluxes in the $2 - 8$~keV band three times higher than sources
that fall outside the wedge. Their X-ray spectra are also harder.  The hardness ratio of the wedge-selected
stack is consistent with moderate intrinsic obscuration, but is not suggestive of a highly obscured, Compton-thick source population.
It is evident from this comparative study that in order to create a complete, unbiased census of supermassive
black hole growth and evolution,  a combination of sensitive infrared, X-ray and hard X-ray 
selection is required. We conclude by discussing what samples will be provided by upcoming survey missions such as {\em WISE}, {\em eROSITA}, and {\em NuSTAR}.

\end{abstract}


\altaffiltext{1}{Space Radiation Laboratory, Mail Stop 290-17, California Institute
of Technology, Pasadena, CA 91125} 
\altaffiltext{2}{NASA Goddard Space Flight Center, Code 662, Greenbelt, MD 20771}
\altaffiltext{3}{NASA Postdoctoral Program (NPP) Fellow}
\altaffiltext{4}{Columbia University, Department of Astronomy, 550 West
120th Street, New York, NY 10027} 
\altaffiltext{5}{Steward Observatory, University of Arizona, 933 North Cherry Avenue, Tucson, AZ 85721}
\altaffiltext{6}{Jet Propulsion Laboratory, California Institute of
Technology, Mail Stop 169-506, Pasadena, CA 91109} 
\altaffiltext{7}{Quest University Canada, 3200 University Blvd., Squamish, BC V8B0N8 CANADA}

\section{Introduction}
\label{sexsiconclusion_sec:intro}

The tight correlation of nuclear black hole mass with the velocity 
dispersion and mass of the galactic
bulge \citep[e.g.,][]{Magorrian:98,Ferrarese:00,Tremaine:02} implies that
the growth and evolution of galaxies is closely linked to the growth
and evolution of the supermassive black holes which reside in (at
least) all massive galaxies \citep[e.g.,][]{Kauffmann:00,Heckman:08}.  
Indeed, recent theoretical work
suggests that feedback from active galactic nuclei (AGN) plays a
dominant role in establishing the present-day appearances of galaxies,
providing a natural, physical explanation for both cosmic downsizing
and the possibly related bimodality in local galaxy properties
\citep[e.g.,][]{Scannapieco:05, Hopkins:08, Cattaneo:09}.  
However, obtaining an unbiased
census of black holes in the universe remains challenging, hampering
our ability to fully probe this connection.  

Most surveys for active
galaxies are severely biased towards unobscured ({\em type~1}) AGN
since nuclear emission in such sources dominates over host galaxy
light at most wavelengths, making type~1 AGN both more readily
identifiable and easier to follow up spectroscopically.  However,
both unified AGN models \citep[e.g.,][]{Antonucci:93,Urry:95} and the shape of the X-ray background suggest
a population of obscured ({\em type~2}) sources which outnumber the type~1 
AGN by up to a factor of ten \citep[e.g.,][]{Comastri:95, Treister:04, Gilli:07}.  
Determining the ratio of unobscured to obscured AGN as a function of 
luminosity and redshift can directly constrain the growth history
of supermassive black holes, and help to quantify their 
influence on the evolution of their host galaxies 
\citep[e.g.,][]{Ueda:03,Barger:05,Hasinger:05,Hopkins:07}. 
Since different search techniques for
obscured AGN suffer from different biases, the problem is best addressed
through multiwavelength studies.

The hard (\hardrange) X-ray and mid-infrared wavebands provide powerful,
complementary methods for identifying and studying AGN over a wide
range of intrinsic obscuration.
Radiation seen from active
galaxies at X-ray energies is primarily due to direct emission from
accretion processes near the central supermassive black hole, with
higher energy photons less susceptible to absorption, while  
mid-infrared
radiation (rest-frame $\lambda_0 \simgt 2~\mu$m) is typically
dominated by emission from dusty obscuring material surrounding
the AGN central engine, 
and is likewise relatively immune to absorption \citep[e.g.,][]{Krolik:99}.
Hard X-ray and mid-infrared surveys of AGN
will therefore sample a wide range of intrinsic absorption
and, in this regard, will be both less biased and more complete than surveys in the
optical \citep[e.g.,][]{Richards:06} and soft X-ray \citep[$E
\simlt 2.4$~keV; e.g.,][]{Hasinger:98, Schmidt:98} bands, both of which
suffer from strong attenuation by even moderate (\nh$\simgt 10^{21}$~cm$^{-2}$) columns of dusty,
obscuring material.  However, AGN samples selected in the
\hardrange\ X-ray and infrared bands will also suffer incompleteness.
Examples exist of AGN identified from optical spectroscopy
which remain undetected even in the deepest X-ray images yet obtained
\citep[e.g.,][]{Steidel:02}; similarly, there are examples of 
optically selected AGN not identified in infrared surveys (e.g., \citealt{Stern:05b}).
Conversely, X-ray missions have
identified AGN whose optical spectra are devoid of AGN signatures
(e.g., X-ray bright, optically normal galaxies, or ``XBONGs'';
\citealt{Comastri:02}).   Therefore a complete census will necessarily
require combining selection techniques from several spectral regimes.

The current generation of X-ray telescopes has dramatically advanced our
knowledge of the AGN population (see~\citealt{Brandt:05}\ for a review).
The {\em Chandra X-ray Observatory} \citep{Weisskopf:96}
provides a large collecting area, a moderate field of view (FOV), and
exquisite angular resolution ($<1$\arcsec) from 0.5 to 8~keV.  These
capabilities have allowed {\em Chandra} extragalactic surveys to
efficiently select and optically identify large samples of AGN in
the 2 -- 8~keV range. Observing in this hard X-ray energy band means
that even sources shrouded by considerable obscuring column densities 
(up to $10^{24}$ cm$^{-2}$) can be detected at low redshift,
and sources with even higher column densities can be detected out
to $z \simgt 2$.  Previous sensitive X-ray telescopes were primarily
restricted to energies below $\sim 2$~keV and thus missed many of
the obscured AGN, a source population that has long been theorized
to explain the mismatch in spectral shape between the \hardrange\
X-ray background ($\Gamma \approx 1.4$) and the unobscured active
galaxies which dominate the source counts at soft X-ray energies
($\Gamma\approx 1.9$; \citealt{Nandra:94}).

The 2003 launch of the {\em Spitzer Space Telescope} \citep{Werner:04}
opened a new era in mid-infrared observations, providing orders of
magnitude improvement in sensitivity in the $3.6-160~\mu$m band.  The increased
sensitivity, combined with the large FOV, allows for the first
time efficient long-wavelength survey capabilities.  The primary
imaging cameras on {\it Spitzer} are the Infrared Array Camera
\citep[IRAC;][]{Fazio:04}, providing simultaneous imaging at 3.6,
4.5, 5.8, and 8~$\mu$m, and the Multiband Imaging Photometer for
{\em Spitzer} \citep[MIPS;][]{Rieke:04}, providing simultaneous
imaging at 24, 70, and 160~$\mu$m.

Several methods have been developed to select AGN based
on their {\em Spitzer} colors~\citep{Lacy:04,Stern:05b,Alonso-Herrero:06}. 
These methods exploit the difference
in the typical spectral energy distribution (SED) of AGN compared
to `normal' galaxies. The near-infrared emission of both typical
galaxies and vigorously star-forming galaxies 
is primarily produced by a thermal stellar population,
resulting in SEDs peaked near the rest-frame 1.6~$\mu$m ``bump.''
This bump, which 
is caused by the minimum in the opacity of the H$^{-}$ ion near 1.6~$\mu$m 
\citep[e.g.,][]{John:88}, 
is a feature of almost all stellar populations \citep[e.g.,][]{Wright:94,Simpson:99},
whereas
AGN-dominated
SEDs have a non-thermal, roughly power-law shape (for $\lambda
\simlt 10~\mu$m). 
At longer wavelengths contributions from stellar blackbody emission
is low, while radiation from the obscuring dust near the central
engine provides strong, relatively isotropic emission which is quite
distinct from stellar light (although the shape of the AGN SED does 
somewhat depend on viewing angle, e.g., \citealt[][]{Elitzur:08}).
\citet{Stern:05b}, for example, suggested a selection technique
exploiting the fact that, for AGN, the long-wavelength side of the
1.6~$\mu$m stellar peak does not decline; the technique uses an
empirically determined `wedge' in IRAC color-color space ([3.6]$-$[4.5]
versus [5.8]$-$[8.0]) that preferentially contains AGN as compared
to normal galaxies or Galactic stars.  Combining a sample of 10,000
$R<21.5$ spectroscopically identified sources from the AGN and
Galaxy Evolution Survey (AGES; Kochanek et al., in prep.) and
mid-infrared observations from the IRAC Shallow Survey
\citep{Eisenhardt:04}, \citet{Stern:05b} defined mid-infrared AGN
selection criteria which robustly identify broad- and narrow-lined
AGN, with only 18\% sample contamination from galaxies (17\%) and
stars (1\%).  The true sample contamination is likely lower, since
many of the spectroscopically normal galaxies may harbor active
nuclei (i.e., are mid-infrared versions of XBONGs).  Working from the
full spectroscopically defined AGES sample, \citet{Stern:05b} found
that the wedge selects 91\% of the broad-lined AGN, 40\% of the
narrow-lined AGN, and fewer than 3\% of the normal galaxies.

This paper presents an exploration of the relative strengths
of {\em Chandra} and {\em Spitzer} as black-hole finders, using a
subset of six fields from the Serendipitous Extragalactic X-ray Source
Identification program (SEXSI; \citealt{Harrison:03}, \citealt{Eckart:05},
and \citealt{Eckart:06}) for which we have obtained mid-infrared coverage with {\em Spitzer}.   
In addition to examining the properties of $\sim 250$ hard X-ray-selected
AGN, we extend our sensitivity to X-ray emission using a
stacking analysis on sources with different mid-infrared characteristics. This enables us to 
make a comprehensive comparison of the
relative characteristics of AGN samples selected from their X-ray and mid-infrared
properties.

We organize the paper as follows:
\S\ref{sexsiconclusion_sec:experiment} introduces the SEXSI program
and presents the complementary mid-infrared observations and the
resulting source catalog; \S\ref{sexsiconclusion_sec:xrayselected}
discusses the mid-infrared properties of the \hardrange\ SEXSI
sources; \S\ref{sexsiconclusion_sec:stacking} presents the mean
X-ray properties of X-ray-non-detected {\em Spitzer} sources;
\S\ref{sexsiconclusion_sec:discussion} provides a discussion; and
\S\ref{sexsiconclusion_sec:conclusions} summarizes our conclusions.
Luminosity calculations assume a standard cosmology used in our 
previous work \citep{Eckart:06},
$\Omega_0 = 0.3$, $\Lambda = 0.7$, and $H_0 = 65~ {\rm km}~ {\rm
s}^{-1}~ {\rm Mpc}^{-1}$.

\section{Observations, Data Reduction and Source Catalog}
\label{sexsiconclusion_sec:experiment}

We have expanded the multiwavelength data available for six SEXSI fields by obtaining {\em Spitzer} imaging observations.  All
six fields have deep {\em Chandra} X-ray images, optical
imaging, and extensive, deep optical spectroscopy --- all of which has been published in
\citet{Harrison:03}, \citet{Eckart:05}, and \citet{Eckart:06}.  In this paper we present {\it Spitzer}
archival and/or targeted observations from 3.6 to 24~$\mu$m from these fields.  Below
we briefly review the SEXSI program, describe the complementary {\em
Spitzer} sample and data reduction, and present the IRAC and MIPS
photometry and source catalog for all \hardrange\ SEXSI X-ray sources.

\subsection{SEXSI X-Ray Sample and Optical Followup}
\label{sexsiconclusion_sec:sexsidata}

The SEXSI survey was designed to obtain a significant sample of
spectroscopically identified objects in the \hardrange\ flux range
from a few times $10^{-13}$ to $10^{-15} \fluxu$, 
to provide a medium-area, medium-depth sample to complement both
the wide-area, shallow surveys (e.g., HELLAS -- \citealt{LaFranca:02};
ASCA Large Sky Survey -- \citealt{Akiyama:00}; ASCA Medium Sensitivity
Survey -- \citealt{Akiyama:03}) and deep, pencil-beam surveys
(e.g., CDF-N -- \citealt{Alexander:03, Barger:03}; CDF-S --
\citealt{Rosati:02, Szokoly:04}).  To accomplish this, SEXSI covers
more than 2 deg$^{2}$ of sky, comprised of twenty-seven archival
{\em Chandra} ACIS images.  The resulting source sample consists
of 1034 hard X-ray-selected sources \citep{Harrison:03} with complete
$R$-band imaging \citep{Eckart:05} and optical
spectroscopy for 477 (46\%) of the sources \citep{Eckart:06}. The
vast majority of these sources are active galaxies with $L_x($\hardrange$)
\simgt 10^{43}\lumin$. Just over half of these AGN have optical
spectra that show typical AGN lines which we classify as either
broad-lined AGN (BLAGN) or narrow-lined AGN (NLAGN).  Extragalactic
sources which exhibit optical spectra devoid of the typical
high-ionization AGN lines are called emission-line galaxies (ELGs)
or absorption-line galaxies (ALGs); such sources are often referred
to as XBONGs in the literature.  See \citet{Eckart:06} for more
details on the SEXSI spectroscopic classifications.

\subsection{{\em Spitzer} Mid-Infrared Imaging}
\label{sexsiconclusion_sec:spitzerreduction}

We obtained mid-infrared imaging for six SEXSI fields through both
archival and targeted {\em Spitzer} programs (see
Table~\ref{sexsiconclusion_tbl:data}).  These programs include
imaging at 3.6, 4.5, 5.8, and 8~$\mu$m from IRAC and imaging at
24~$\mu$m from MIPS.  Fields were chosen both to optimize use of
archival {\it Spitzer} data and to target SEXSI fields with the
most complete spectroscopic coverage.  Program ID~20808 was designed
specifically for the SEXSI follow-up, with typical exposure times
of 600~s for IRAC observations and 1000~s for MIPS observations.
Because the SEXSI data were drawn from the {\em Chandra} public
archive, the fields are not `blank': the target sources in the six
SEXSI {\em Spitzer} fields are all $z > 0.18$ galaxy clusters,
five of which have $0.64 < z < 1.3$
(see Table~\ref{sexsiconclusion_tbl:data}).

\subsubsection{IRAC Observations and Reductions}

Our IRAC data reduction begins with the basic calibrated data (BCD)
output from the {\it Spitzer} Science Center pipeline (ver.~10.0.1).
Note that {\it Spitzer} BCD images have surface brightness units
in MJy sr$^{-1}$, where 1~MJy sr$^{-1}$ is equal to 23.5~$\mu$Jy
arcsec$^{-2}$.  Most of the IRAC observations consist of six
dithered 100~s exposures. For each IRAC channel, the BCDs were
mosaicked using the MOPEX package \citep{Makovoz:05} and resampled
by a factor of two to give a pixel scale of 0.61\arcsec.  The MOPEX
outlier (e.g., cosmic ray, bad pixel) rejection was optimized for
the regions of deepest coverage in the centers of the images.  Source
extraction was performed with SExtractor \citep{Bertin:96}\ both
in single-image mode to produce independent source catalogs in all
four IRAC bands and in dual-image mode using the 3.6~$\mu$m channel
for source detection.  All IRAC BCD images have astrometry derived
from the 2MASS survey and are good to better than $0.5$\arcsec.

In general, we use a 5\arcsec\ diameter aperture for IRAC photometry.
For the small fraction of the sources ($\sim 5$\%) that are spatially
resolved and larger than this, we use a 7\arcsec\ diameter aperture.
We select the apertures to ensure a fixed extraction
area similar to the isophotal area; this same methodology was adopted
by both the {\em Spitzer} First Look Survey \citep[FLS;][]{Lacy:05}
and the {\em Spitzer} High-Redshift Radio Galaxy program
\citep[SHzRG;][]{Seymour:07}.  The aperture corrections, applied
after the conversion from surface brightness to flux density, are
1.167, 1.213, 1.237 and 1.466 in the four IRAC bands, respectively,
for the 5\arcsec\ apertures.  The corresponding corrections
for the 7\arcsec\ apertures are smaller (1.100, 1.130,
1.109 and 1.198, respectively).  We derive total uncertainties on
the resulting IRAC flux densities by adding in quadrature the
statistical uncertainties (derived from SExtractor) to a conservative
10\%\ estimate of the systematic uncertainties.  The latter
includes several effects \citep[e.g.,][]{Hora:08}: uncertainties in the IRAC zero point, field-dependent
color corrections, pixel phase effects (e.g., zero point corrections
that depend on the position of a source relative to the nearest
pixel center), and variation of the pixel scale across the field
of view. For nondetections we calculate $5 \sigma$ upper limits.
A small number of sources are recovered by SExtractor with signal-to-noise ratios 
below 5 ($3<$~S/N~$<5$); 
examination of the images shows that these are valid detections, 
generally near the edge of the mosaic.  We flag these sources in the 
catalog and exclude them from the analysis presented here. 

\subsubsection{MIPS Observations and Reductions}

MIPS observations obtained for this program (e.g., PID 20808; see
Table~\ref{sexsiconclusion_tbl:data}) consist of two cycles of 30~s exposures with a $3 \times
3$ dither pattern at 24~$\mu$m.  This observing strategy provides
an average (median) exposure time of 1290~s (877~s) across our
combined fields of view.  To process the MIPS data, the BCDs provided
by the {\em Spitzer} Science Center were first median filtered to
remove a stripe pattern artifact in the individual frames.
We  then mosaicked the data with MOPEX using standard inputs to
produce final, stacked images with the native platescale of
2.45\arcsec\ pix$^{-1}$.  We convert the mosaics  from surface
brightness units to flux density units, and extract flux densities 
with SExtractor using an 11.76\arcsec\ diameter extraction
aperture with an aperture correction of 1.697.  
Calibration with MIPS is performed with stars using a blackbody SED.  
\citet{Fadda:06}\ found that the color corrections for both starburst 
galaxies and AGN were small at $z<1$.  At $z>1$, the color correction for a 
prototypical luminous AGN (NGC 1068) remains small.  As most SEXSI 
sources fall within one of these categories, we did not apply a color 
correction to account for MIPS sources having non-stellar SEDs.
As with the IRAC photometry, we derive total uncertainties by adding in quadrature
the statistical uncertainties (derived from SExtractor) to a
conservative 10\%\ estimate of the systematic uncertainties 
\citep[see, e.g.,][]{Engelbracht:07};
for the MIPS data the statistical uncertainties are generally the dominant 
source of error.
Compared with the IRAC photometry, a larger number of MIPS sources are
detected with signal-to-noise ratios below 5 ($3<$~S/N~$<5$). 
We flag these sources in the catalog and 
exclude them from our analysis. For the true non-detections we calculate
$5\sigma$ upper limits.

\subsection{Source Catalog}
\label{sexsiconclusion_sec:catalog}

Table~\ref{sexsiconclusion_tbl:sexsispitzercat}\ presents mid-infrared
photometry for 290 hard X-ray-selected SEXSI sources.  The full
machine-readable catalog is available online.  Each of the four
IRAC catalogs as well as the MIPS catalog were individually matched
to the SEXSI X-ray source positions\footnote{We use the X-ray
positions from \citet{Eckart:05} which have been corrected for
small {\em Chandra} pointing offsets.} using a 2.5\arcsec\ search
radius.  To calculate a false match rate we shifted the X-ray source
catalog by 1\arcmin\ and matched to the IRAC and MIPS catalogs; this entire
procedure was repeated six times using different 1\arcmin\ shifts. 
The resulting false 
match rates are: 10.1\% ($3.6~\mu$m), 
7.2\% ($4.5~\mu$m), 3.7\% ($5.8~\mu$m), 2.6\% ($8.0~\mu$m), 1\% ($24~\mu$m), 
and $<1$\% for four-band-detected IRAC sources. 
For ease of reference, columns (4)--(10) of 
Table~\ref{sexsiconclusion_tbl:sexsispitzercat}\ present X-ray and
optical source information from previous SEXSI publications, while
the {\em Spitzer} mid-infrared data is presented in columns (11)--(15).
Complete X-ray source information is provided in \citet{Harrison:03}
and optical-followup source catalogs are presented in \citet{Eckart:05}
and \citet{Eckart:06}.

Column (1) presents the source ID, followed by $\Delta \alpha$ and
$\Delta \delta$, the difference in right ascension and declination
between the X-ray position and the {\em Spitzer} counterpart.  Column
(4) lists the \hardrange\ flux in units of $10^{-15} \fluxu$,
while column (5) gives the X-ray hardness ratio, $HR \equiv (H-S)/(H+S)$, 
where $H$ is the hard-band counts ($2-10$~keV) and $S$ is the 
soft-band counts ($0.5-2$~keV).
The optical-counterpart $R$-band magnitude is presented in column
(6), followed by the optical-spectroscopic classification (7)
and the source redshift (8). Sources marked `unid' have no
spectroscopic classification or redshift.  The observed X-ray
luminosity is given in column (9) and the \nh\ value determined
from the X-ray data is found in column (10). 
When the best-fit \nh\ measurement is zero, we report ``$<$'' in the table;
see \citet{Eckart:06}\
for details of these calculations.  Columns (11)--(14) provide the
flux density in each of the IRAC bands with associated errors. The errors
are $\simgt 10$\% of the source flux owing to our conservative
estimate of systematic uncertainties for the {\em Spitzer} data.
The MIPS data are presented in the final column.  Entries in Columns
(11)--(15) that are blank indicate that a source was not observed
in that band.

\section{Mid-Infrared Properties of X-ray Selected AGN}
\label{sexsiconclusion_sec:xrayselected}

This section explores the mid-infrared properties of the sample 
of \hardrange\ SEXSI sources. In particular, we focus on whether
the X-ray-selected AGN will also be identified as AGN via the 
\citet{Stern:05b}\ IRAC color-color AGN selection technique, 
and how this mid-infrared identification depends on various source
properties, such as optical spectroscopic classification, redshift, 
X-ray flux and mid-infrared flux density (\S \ref{sexsiconclusion_sec:flux}), X-ray luminosity 
(\S \ref{sexsiconclusion_sec:luminosity}), and $R$-band magnitude 
(\S \ref{sexsiconclusion_sec:rmag}).  Section~\ref{sexsiconclusion_sec:compare}\ 
provides a comparison to AGN samples presented in the literature: 
\S \ref{sexsiconclusion_sec:compare_hardxray}\ compares the mid-infrared
properties of hard X-ray-selected samples, while \S \ref{sexsiconclusion_sec:type2quasar}\ 
discusses the identification of high-redshift, obscured quasars via $3.6~\mu$m$-24~\mu$m 
selection criteria.

The {\em Spitzer} data covers 290 hard X-ray SEXSI sources, of which
154 have spectroscopic redshifts.  However, a significant fraction
of the X-ray sources do not have coverage in all four IRAC 
bands.\footnote{IRAC has two adjacent fields of view that are imaged
in pairs (3.6 and 5.8~$\mu$m; 4.5 and 8.0~$\mu$m).
The sources in our sample that lack four-band coverage typically lie nearer to the edge
of the X-ray survey area and were imaged by only one IRAC FOV, 
thus yielding only two-band coverage.}
For the following analysis we focus on the 189 hard X-ray
sources (65\%) with observations in all four IRAC bands.  For some analyses,
we consider the slightly smaller sample of 155 sources (53\%) 
with MIPS 24~$\mu$m observations as well.  Because of the relative depths
of the mid-infrared and X-ray data, nearly all
of the SEXSI sources have infrared counterparts in one or more of
the bands.  Overall, $98\%$ of SEXSI sources observed at $3.6~\mu$m are
detected at greater than $5 \sigma$ at that wavelength, and the
detection rates for the other IRAC bands are all $\simgt 90$\%.  
Of the four-band IRAC-observed
SEXSI sources, 57\% have optical spectroscopic redshifts, and, of
those, 89\% are detected in all four IRAC bands.  The corresponding
MIPS detection rate is 54\% (68\% if $<5\sigma$ detections are included).

\begin{figure*}
\centering
\includegraphics[width=7in]{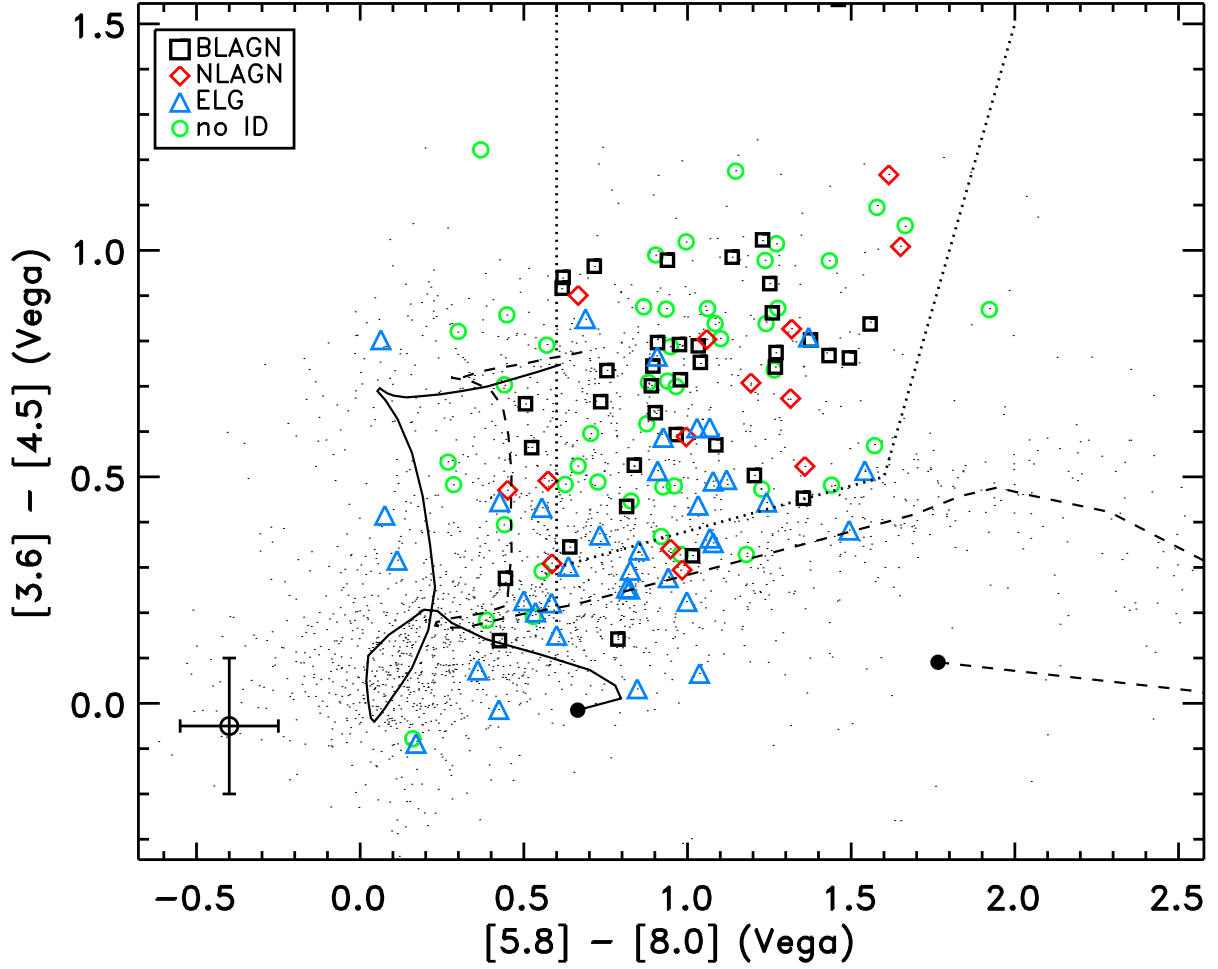}
\caption{Color-color diagram of sources detected at $\geq 5 \sigma$
in all four IRAC bands (dots).  Sources with a \hardrange~X-ray
detection have a larger symbol overlaid, coded based on the optical
spectroscopic classification (see inset).  The dotted lines demarcate
the AGN selection wedge introduced by \citet{Stern:05b}. The $0\le
z \le 3$ color tracks for two non-evolving galaxy templates from
\citet{Devriendt:99}\ are illustrated; the large filled circles
indicate $z=0$. A starburst galaxy is illustrated with the track
of M82 (dashed line) and NGC 4429, an S0/Sa galaxy with a star-formation
rate $\sim 4000 \times$ lower, is indicated with the solid line. 
Median errors are illustrated at the lower left.}

\label{sexsiconclusion_fig:wedgeplot}
\end{figure*}

Figure~\ref{sexsiconclusion_fig:wedgeplot} shows an IRAC color-color
diagram of all mid-infrared sources with $\geq 5 \sigma$ detections
in all four IRAC bands (small black dots).  Larger symbols indicate
SEXSI \hardrange\ sources, as detailed in the caption.  The dotted
lines indicate the \citet{Stern:05b}\ AGN selection wedge. We 
adopt the \citet{Stern:05b}\ selection, which has slightly less contamination 
than the criteria suggested by \citet{Lacy:04} based on [4.5]$-$[8.0]
versus [3.6]$-$[5.8] color space \citep[e.g.,][]{Donley:08}. 
The majority (63\%, or 90/142) of the X-ray-selected AGN lie in the
\citet{Stern:05b}\ selection wedge.  Considering the spectroscopic
sample, the AGN wedge contains 85\% (34/40) of the BLAGN, 64\%
(9/14) of NLAGN, and 34\% (13/38) of the ELG, as shown in
Figure~\ref{sexsiconclusion_fig:wedgeplot}.  Since there were only
two ALGs (and one Galactic star) in the SEXSI sample with four-band IRAC observations, 
these minor source populations are not considered further.  Including 
1$\sigma$ errors on the IRAC flux densities, all but three of the BLAGN are
consistent with the wedge (one of these is the lowest-luminosity
BLAGN in the sample, with $L_{\rm 2-10~keV}  = 4 \times 10^{42} \lumin$) and all but one of the NLAGN
are consistent with the wedge selection.  In the following sections
we explore the AGN selection further; in particular, we discuss the
discrepancy between the high fraction of BLAGN and NLAGN selected
using the mid-infrared techniques as compared to low fraction of
ELGs.\footnote{Ten percent of the spectroscopically identified 
X-ray-selected AGN lie at redshifts near that of the target cluster
of the {\em Chandra} observations.
Throughout \S \ref{sexsiconclusion_sec:xrayselected}\ we
include these sources in our analysis; however, we note that our 
conclusions remain unchanged if this 
ten percent of sources is eliminated from the sample.}

\subsection{Dependence on Observed X-Ray Flux and Mid-Infrared Flux Density}
\label{sexsiconclusion_sec:flux}

\begin{figure*}
\centering
\includegraphics[width=7in]{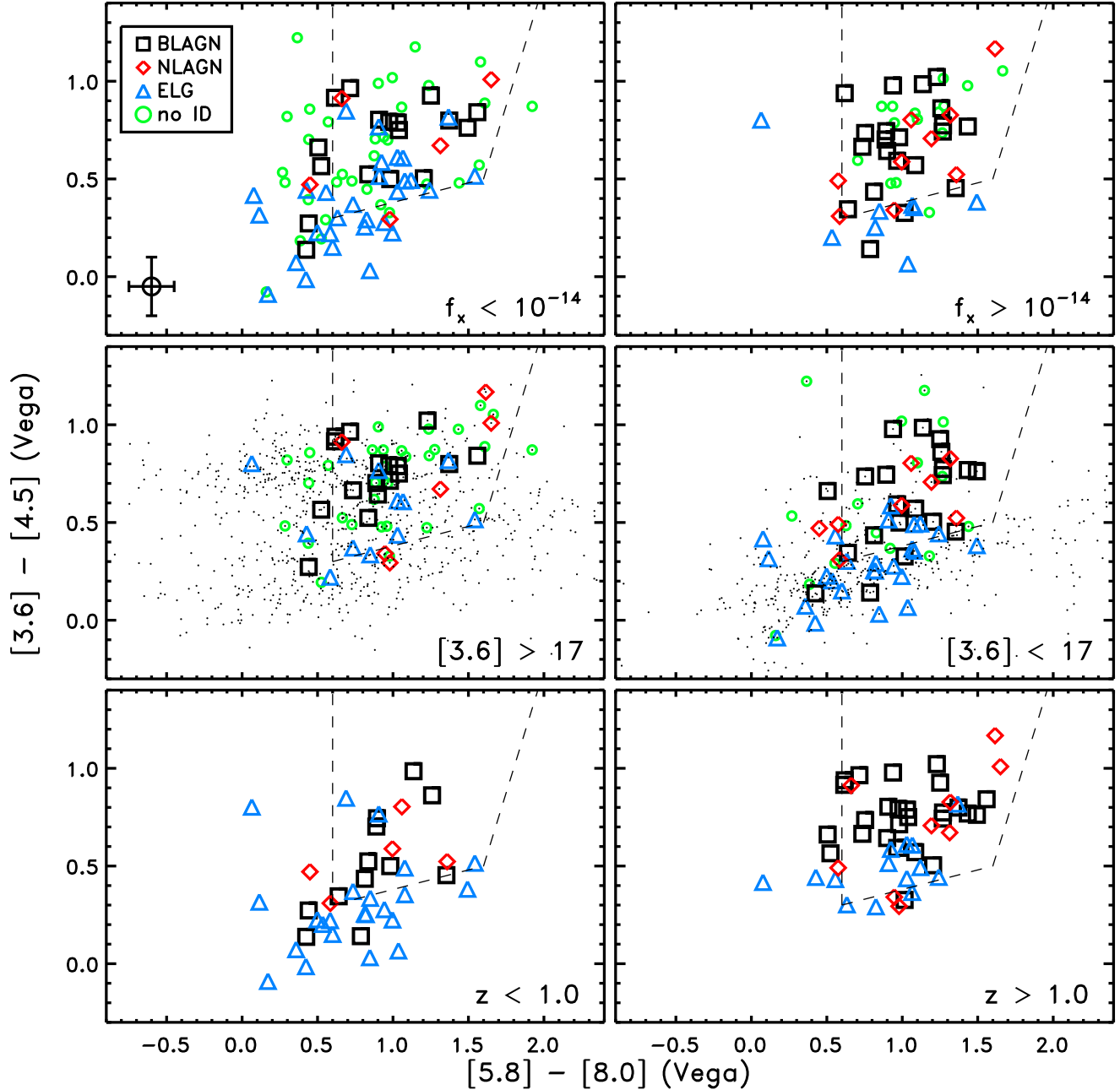}
\caption{ Mid-infrared wedge selection for samples divided by hard
X-ray flux (upper panels; units of erg~ cm$^{-2}$~ s$^{-1}$), 3.6
$\mu$m flux density (middle panels; Vega magnitudes) and 
redshift (lower panels). The upper left panel shows a median 
error bar for each color. Wedge selection is much more efficient at higher X-ray fluxes, as
the brighter X-ray sources are predominantly BLAGN.  However, wedge
selection is not more efficient at higher mid-infrared flux densities; a
significant number of SEXSI ELG AGN outside of the wedge are brighter
than [3.6] = 17 (44~$\mu$Jy).  The background dots in the middle
panel show all four-band IRAC sources in the SEXSI/{\em Spitzer}
data; comparison to Figure~\ref{sexsiconclusion_fig:wedgeplot} shows
that the brighter sources have IRAC colors similar to starburst
galaxies.  The lower panel shows that the sources missed by the
wedge are mainly at low redshift.  Taken together, these results
imply that shallow mid-infrared surveys would detect a large number
of AGN that have galaxy colors in the infrared, do not show strong
AGN lines in their optical spectra, and are X-ray faint.  These
sources can only be readily identified with deep X-ray surveys.
}
\label{sexsiconclusion_fig:wedge_cuts}
\end{figure*}

Mid-infrared selection of AGN relies exclusively on observed
mid-infrared colors to distinguish AGN from galaxies.
This selection can be compared to X-ray identification of AGN
for various flux limits in both wavelength regimes.
Figure~\ref{sexsiconclusion_fig:wedge_cuts} explores the wedge 
membership of our sample for flux cuts
in the X-ray and mid-infrared.  The top panels show that 
X-ray-bright AGN are predominantly
found in the wedge, while for fainter X-ray sources the wedge
membership fraction decreases.  However, the opposite is true when
considering the mid-infrared data:  a significant fraction of {\em
brighter} X-ray-selected AGN are found outside the wedge.  Roughly
half of the SEXSI AGN brighter than $f_{3.6 \mu{\rm m}} = 50~\mu$Jy
([3.6]$< 16.86$) are outside the wedge, compared to one-third for
fainter mid-infrared sources.  The non-wedge X-ray AGN are also
generally at low redshift.

Figure~\ref{sexsiconclusion_fig:wedge_cuts} (middle panels) also shows X-ray
undetected mid-infrared sources.  At bright mid-infrared fluxes,
most of the IRAC sources have colors consistent with
low-redshift galaxies (e.g., see galaxy tracks in Figure~\ref{sexsiconclusion_fig:wedgeplot}).
The X-ray sources outside the wedge have similar colors, suggesting
that their mid-infrared flux is dominated by stellar emission as
opposed to AGN activity.  Many of these non-wedge-selected X-ray sources indeed have optical
spectra of ELGs, consistent with the idea that emission related to star
formation outshines the emission from the active nucleus in some 
parts of the spectrum (i.e., in the mid-infrared and optical); however, 
the X-ray luminosities of these sources are too high to 
be attributed wholly to emission related to star formation and instead
we attribute the X-ray flux to emission from the nuclei.
Figure~\ref{sexsiconclusion_fig:wedge_cuts}
also shows that while a large fraction of the brightest mid-infrared
IRAC AGN candidates are detected in the SEXSI {\em Chandra} images,
this is not the case at fainter mid-infrared magnitudes:  there is
a large population of mid-infrared-selected AGN candidates which
lack X-ray detections, 
even in our relatively sensitive ($\langle t_{\rm
exp} \rangle \approx 100$~ks) {\em Chandra} data sets.

Figure~\ref{sexsiconclusion_fig:f36vsfx} shows the mid-infrared and
X-ray flux distributions of SEXSI sources, and gives an indication
of the efficiency of mid-infrared AGN selection as a function of
the imaging depth in both wavelength regimes.  The six fields considered
here cover a relatively small area of the sky, and thus there are few bright
sources in the sample.  However, despite these small numbers,
our finding that a relatively small proportion of SEXSI AGN with bright mid-infrared 
fluxes are selected by the mid-infrared wedge shows that shallow infrared
surveys will not be sensitive to the same AGN population as X-ray surveys.
We consider these results in light of
upcoming shallow, full-sky survey missions such as {\em eROSITA}
and the {\em Wide-field Infrared Survey Explorer} {\em (WISE)} in
the our concluding section (\S \ref{sexsiconclusion_sec:finalconclusions}).

\subsection{Dependence on X-ray and Infrared Luminosity}
\label{sexsiconclusion_sec:luminosity}

\begin{figure*}
\centering
\includegraphics[width=7in]{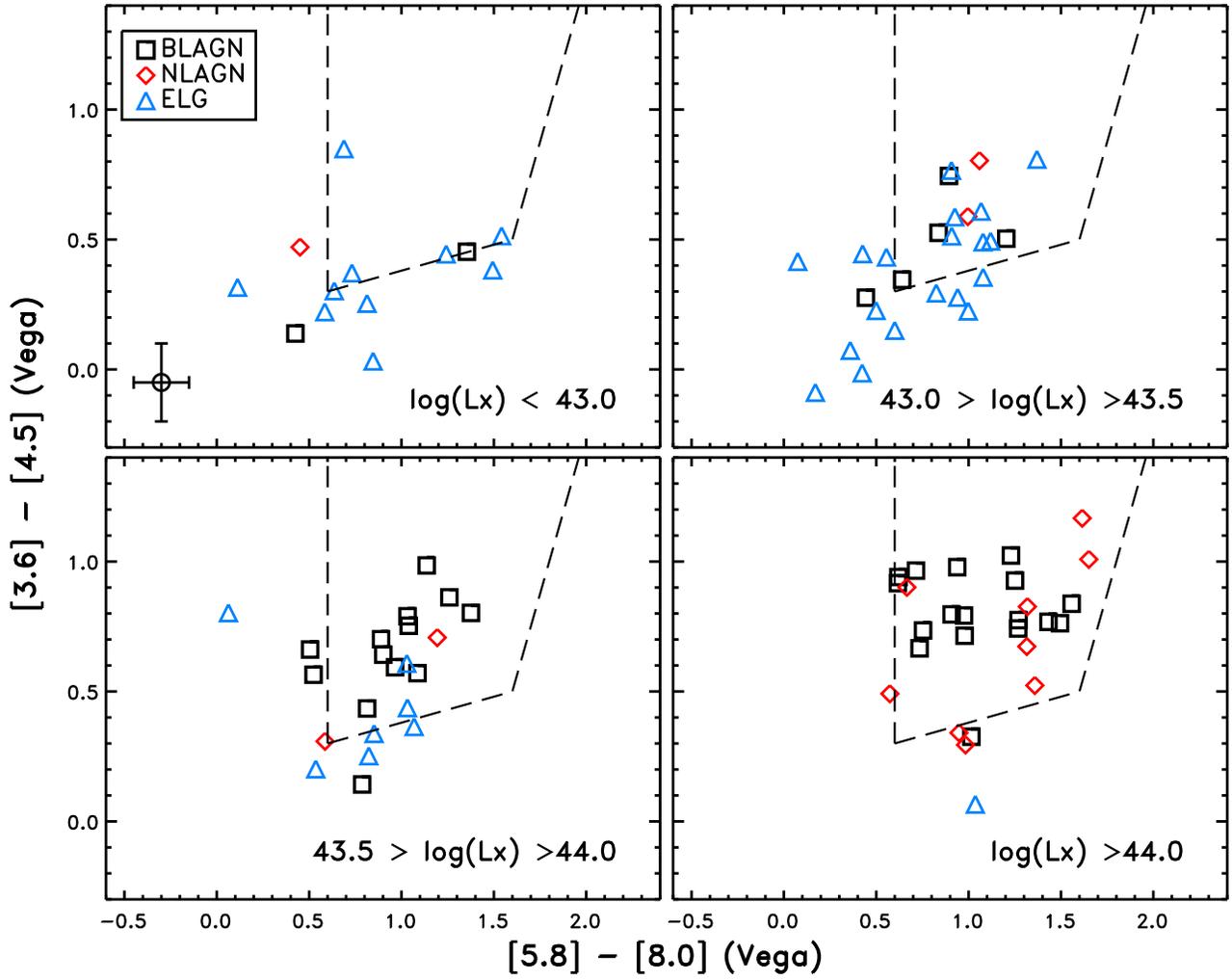}
\caption{IRAC color-color diagram of SEXSI \hardrange\ sources,
split by absorption-corrected intrinsic X-ray luminosity.  As the
X-ray luminosity grows, a higher fraction of sources fall solidly
into the \citet{Stern:05b} mid-infrared AGN selection wedge. }

\label{sexsiconclusion_fig:wedgelum}
\end{figure*}

Figure~\ref{sexsiconclusion_fig:wedgelum}\ illustrates the wedge
selection for four \hardrange\ luminosity ranges, showing that X-ray
luminosity strongly affects the fraction of sources which appear
as AGN in the mid-infrared.  For this analysis, we use the rest-frame 
\hardrange\ X-ray luminosities
from \citet{Eckart:06} that were derived assuming $\Gamma =
1.5$.  The entire lowest luminosity subsample ($L_x<10^{43}\lumin$)
lie either outside of the wedge or near the outskirts of the wedge,
while the fraction within the wedge increases monotonically with
X-ray luminosity.  At the higher X-ray luminosities ($L_x>10^{43.5}
\lumin$), all but two of the sources are consistent with the wedge.
Of the 36 sources with spectroscopic identifications that are not selected
by the wedge, 25 are classified as ELGs based on their optical
spectra. These sources have a mean redshift of $\langle z \rangle
\sim 0.8$, average infrared luminosity $\langle L_{4.5} \rangle
\sim 10^{44}~\lumin$, and average X-ray luminosity $\langle L_x
\rangle \sim 10^{43.2}~\lumin$.  Table~\ref{sexsiconclusion_tbl:wedgelum}\
tabulates the fraction of X-ray sources in the wedge for the four
plotted X-ray luminosity ranges.  The X-ray luminosities used for
binning the sources are the intrinsic, absorption-corrected
luminosities described in \citet{Eckart:06}; 
these results are not significantly different, however, if
observed (e.g., uncorrected) luminosities are used instead.

Overall, it is clear that high X-ray luminosity sources
predominantly show strong AGN features in their optical spectra and
are wedge-selected, while the low-X-ray-luminosity sources generally lack
AGN spectroscopic features and are not in the wedge.  These results
show that \hardrange\ X-ray surveys are efficient at finding
low-luminosity AGN, sources missed by mid-infrared selection
techniques.
This same trend is seen when the sample is divided by mid-infrared
luminosity: sources with lower mid-infrared luminosities tend to
fall out of the wedge.\footnote{We do not apply K-corrections when calculating the 
mid-infrared luminosities as they are not well known. We estimate that for 
quasar-like spectra the K-correction will not be important, while for 
galaxy-like spectra it will vary a factor of $\approx 4-5$ between $z=0.5$ and 
$z=1.5$. Overall, these K-corrections will not change the fact that $L_x$ and $L_{IR}$ are
correlated, however there will be scatter based on the particular SED.} 
This is unsurprising given that, in luminous
AGN, nuclear emission reprocessed by dust dominates the mid-infrared
continuum and thus the mid-infrared luminosity is strongly tied to
the AGN bolometric output.

\subsection{Dependence on $R$-band Magnitude}
\label{sexsiconclusion_sec:rmag}

\begin{figure}[t]
\centering
\includegraphics[width=6in]{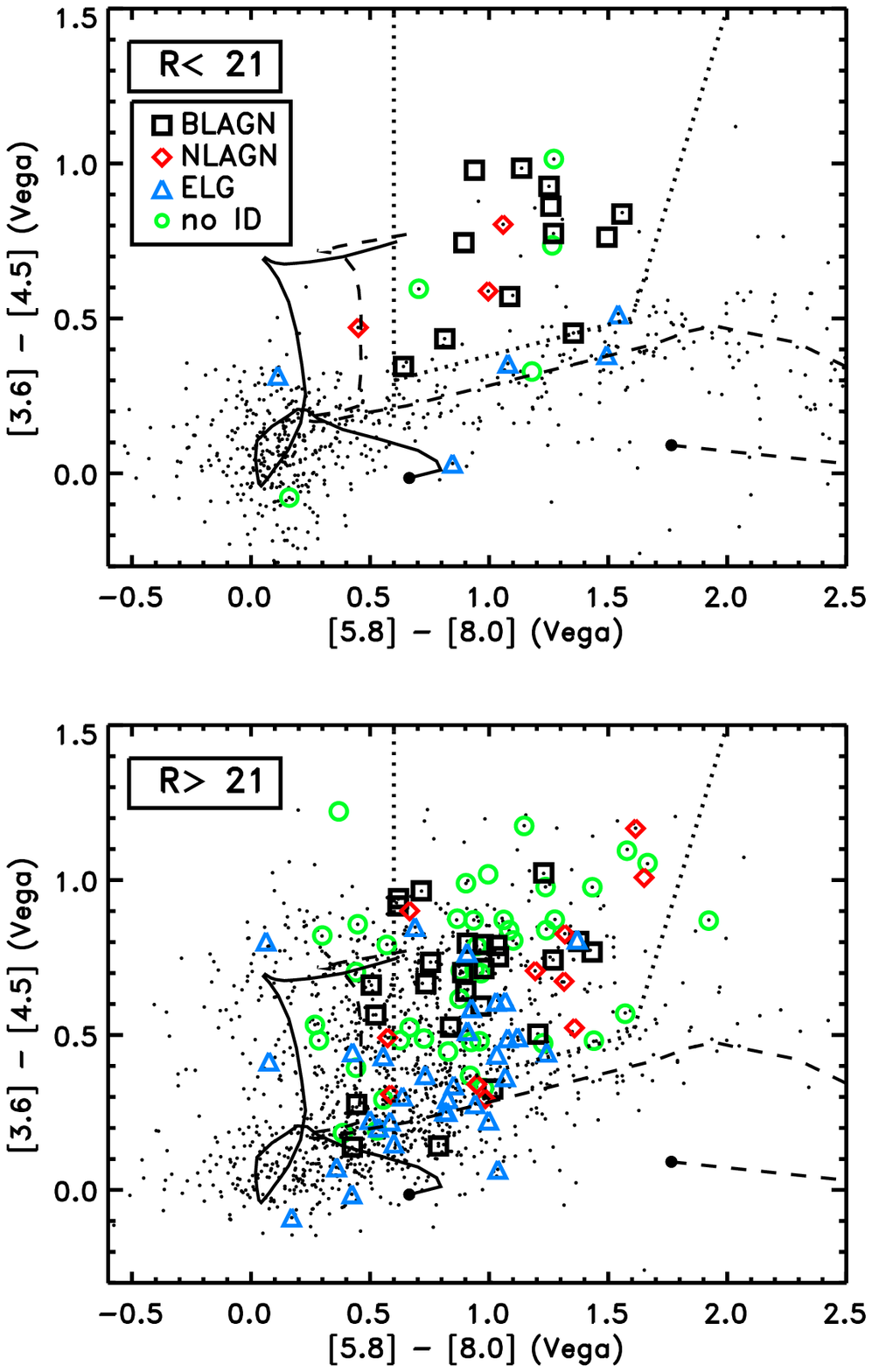}
\caption{IRAC color-color diagram of sources with optical counterparts
brighter (top panel) and fainter (bottom panel) than $R=21$. The
galaxy tracks are the same as in
Figure~\ref{sexsiconclusion_fig:wedgeplot}.}
\label{sexsiconclusion_fig:wedgeplot_rcuts}
\end{figure}

Figure~\ref{sexsiconclusion_fig:wedgeplot_rcuts}\ shows this same
color-color diagram as in Figures~\ref{sexsiconclusion_fig:wedgeplot}, 
\ref{sexsiconclusion_fig:wedge_cuts}, and \ref{sexsiconclusion_fig:wedgelum}\
but split at $R_{\rm cut}=21$; the top panel
shows {\em Spitzer} and SEXSI sources with bright optical counterparts
and the bottom panel shows those with counterparts fainter than $R_{\rm cut}=21$.
Tabulated results comparing the wedge selection shown in these plots
are presented in Table~\ref{sexsiconclusion_tbl:wedgermag}.  These
data show a few strong trends.  Only 5\% $\pm~1$\% of optically bright four-band
IRAC sources lie in the wedge, while 29\% $\pm~2$\% of the optically faint
four-band IRAC sources do.  If we consider the subset of these
sources that have \hardrange\ detections, we see that the 70\% $\pm~16$\% of
the optically bright X-ray sources lie in the wedge and 61\% $\pm~7$\% of the
optically faint X-ray sources are in the wedge. The X-ray sources
in general are more likely to fall in the wedge than a typical IRAC
source, consistent with the notion that the wedge preferentially
identifies AGN \citep[e.g.,][]{Gorjian:08, Ashby:09}.  An interesting
trend evident from this sample is the large fraction of $R>21$ X-ray
undetected wedge sources: of the 374 IRAC sources in the wedge
with $R>21$, only 18\% $\pm~2$\% (68 sources) are X-ray detected. 
For comparison, the
corresponding optically bright ($R<21$) sources are detected by
{\em Chandra} nearly two-thirds (63\% $\pm~14$\%) of the time.  This trend is consistent
with the idea that many of the X-ray undetected {\em Spitzer} AGN
lie at high redshift and/or are heavily obscured, resulting in faint 
optical counterparts and X-ray non-detections.  The infrared data
is deep compared to the X-ray data, which, in combination with the 
scatter in the X-ray to infrared flux ratios, means that
sources that lie at high redshift may fall below the X-ray detection
threshold yet have significant detections in the {\em Spitzer} data. 
Sources at moderate redshift with unobscured
\hardrange\ fluxes that would easily be detected in our {\em Chandra} data
will fall below the detection 
threshold of even the deepest {\em Chandra} coverage in this study ($\sim 200$~ks, limiting flux of $\approx 2 \times 10^{-15}~\fluxu$)
if they are heavily obscured. For example, the observed flux of a source at $z= 0.4$ with 
an unobscured \hardrange\ flux of $f_x \approx 1\times 10^{-14}~\fluxu$ will 
be reduced to $f_x= 2\times 10^{-15}\fluxu$ by an obscuring column of $1 \times 10^{24}$~cm$^{-2}$, 
assuming photo-electric absorption and $\Gamma=1.9$.  
We also note that a significant number of the
fainter wedge sources may also be normal galaxies 
\citep[see also, e.g.,][]{Donley:08}; the typical
600~s IRAC integrations considered here reach significantly ($\approx
3\times$) deeper than the shallow data sets from which the original
wedge criteria were derived.

The final rows of Table~\ref{sexsiconclusion_tbl:wedgermag}\ split
the population by spectroscopic source type; the columns for {\em
Spitzer} sources are blank here because we do not have spectroscopic
followup of the non-X-ray-detected {\em Spitzer} AGN candidates,
though follow-up of optically-faint, mid-infrared-selected AGN from
the IRAC Shallow Survey shows a large fraction of them to be type~2
AGN at $z \simgt 2$ (Stern et al., in prep.).  Of the optically
bright ELGs, only one of six sources lies within the wedge (and
it is at the boundary), while twelve of thirty-two $R>21$ ELGs
lie in the wedge.  This trend is different from that seen with the
BLAGN and NLAGN, where a similar or higher fraction
of $R<21$ sources lie in the wedge. This may
indicate that for optically bright ELGs the starburst component of
the SED is brighter than the AGN component pushing the source away
from the AGN-defined wedge, while for some of the optically fainter
sources the AGN component is strong in the mid-infrared leading to
wedge selection. This trend is explained by higher X-ray luminosity
sources tending to fall in the wedge: as ELGs are identified at
higher redshift they will be fainter in $R$ and also have higher
average $L_x$ due to selection effects, and thus preferentially
fall in the wedge as compared to low-luminosity, low-redshift
sources. Splitting the wedge sources by redshift confirms this
assertion (Figure~\ref{sexsiconclusion_fig:wedge_cuts}).  
Overall, regardless of $R$-magnitude, the
majority of the ELGs fall outside of the wedge.

\subsection{A Brief Summary of AGN Selection in the X-ray and Mid-Infrared}

The results of our comparison of AGN selection techniques presented in this
section may be summarized as follows:

\begin{itemize}

\item{nearly two-thirds (63\%) of X-ray-selected AGN lie in the mid-infrared selection wedge
and
nearly all of the X-ray-selected AGN that have optical spectra that exhibit high-ionization
emission lines (i.e., BLAGN and NLAGN) lie within the mid-infrared selection wedge, within 
photometric uncertainties;}

\item{on average, the X-ray-selected AGN classified as ELGs tend
to fall outside of the infrared selection wedge, and thus are sources that would be 
missed by both mid-infrared and optical AGN selection techniques;}

\item{X-ray bright sources tend to fall in the AGN selection wedge, while sources
with lower X-ray fluxes tend to fall outside the wedge;} 

\item{the opposite is true when 
considering mid-infrared fluxes: the
X-ray-selected AGN that are faint in the mid-infrared tend to fall in the wedge, 
while those that are bright in the
mid-infrared tend to be missed by the infrared AGN selection technique;}
 
\item{the non-wedge X-ray-selected AGN generally lie at low redshift;}

\item{overall, high X-ray luminosity sources fall within the mid-infrared selection wedge, 
while the low-luminosity sources fall outside the wedge: the BLAGN and NLAGN are predominantly wedge selected, as they have higher X-ray luminosities, while the ELGs, which have lower X-ray luminosities, tend 
to fall outside the selection wedge;}

\item{there is a large population of mid-infrared, {\em Spitzer}-selected AGN that 
is not detected in the \hardrange\ X-ray data;}

\item{when mid-infrared-selected AGN sources are split into
optically bright ($R<21$) and optically faint ($R>21$) populations, the
optically bright sources are detected by {\em Chandra} the majority of the time, 
while only $\approx 18$\% of the optically faint sources are identified in the X-ray.}

\end{itemize}

\subsection{Comparison to Previous Work}
\label{sexsiconclusion_sec:compare}

\subsubsection{Hard X-ray Selection}
\label{sexsiconclusion_sec:compare_hardxray}

We compare the SEXSI {\em Spitzer} sample to several other {\em
Chandra} -- {\em Spitzer} surveys that have a range of depth and area.
The {\em XMM-Newton} Medium Deep Survey \citep[XMDS;][]{Pierre:07}
identified $2-10$~keV X-ray sources over 1 deg$^2$ with
mid-infrared coverage from the {\em Spitzer} Wide-Area Infrared
Extragalactic Survey survey \citep[SWIRE;][]{Lonsdale:03}.
\citet{Polletta:07} and \citet{Tajer:07} present a catalog of 122
X-ray-selected AGN with $f_x \simgt 10^{-14}~\fluxu$,  about a factor five
shallower than the SEXSI sample considered here.  All of these
sources have at least two-band IRAC detections.  Using the
\citet{Lacy:04} mid-infrared wedge selection technique, they find
that $80-95$\% of the X-ray sample would be selected on the basis
of mid-infrared colors.  We find that 82\% of the XMDS AGN are selected by the
Stern wedge, much higher than the 62\% of SEXSI sources within the
wedge. If a flux cut of $f_x >10^{-14}~\fluxu$ is applied to the
SEXSI sources in order to match the limit of XMDS, the wedge
membership increases to 74\%; when scatter due to photometric errors
is considered, the wedge membership of the X-ray-brighter SEXSI
sources is similar to XMDS 
(see upper panel of Figure ~\ref{sexsiconclusion_fig:wedge_cuts}).

\citet{Barmby:06} present {\em Spitzer} identifications for a catalog
of $\sim150$ X-ray AGN in the Extended Groth Strip. The sample
includes 138 sources detected in a 200~ks {\em Chandra} exposure.
Of these, 67 would be selected on the basis of hard-band detection
and thus form a comparable sample to SEXSI. For the full catalog,
\citet{Barmby:06} report that only 50\% of sources are within the
Stern wedge. Considering the hard-band selected sample, this
percentage increases to 57\%. If a further cut is made to select
sources with $f_x>2\times10^{-15}~\fluxu$ (similar to the typical
SEXSI limit), the percentage of wedge sources increases to 64\%,
and for a flux cut of $f_x>1\times10^{-14}~\fluxu$ (similar to XMDS)
the percentage reaches 71\%. These trends show that various 
surveys agree when similar selection criteria are applied.

Our results are also consistent with those of \citet{Donley:07},
who use deeper X-ray data over a smaller area from the Ms~{\em
Chandra} Deep Field-North \citep{Alexander:03} and find that a small
fraction of low-$L_x$ sources fit their mid-infrared power-law AGN
selection criterion, with only one-third being within the Stern wedge.
The mid-infrared selection efficiency for their sample increases 
from $\sim 14$\% at $L_{\rm 0.5-8~keV}<10^{42} \lumin$ to 
100\% at $L_{\rm 0.5-8~keV} > 10^{44} \lumin$.

Combining these results, a consistent picture emerges in which hard
X-ray surveys uncover a population of low-luminosity AGN missed by
mid-infrared selection techniques, with the efficiency of mid-infrared
selection declining in proportion to the depth of the X-ray data.
Optical selection techniques which rely on the detection of
high-ionization narrow emission lines in order to identify obscured
AGN \citep[e.g.,][]{Zakamska:03}  similarly miss this population
of low-luminosity AGN.

Lastly, we can compare to the large sample of infrared-selected AGN
assembled by \citet{Hickox:07} from the Bo\"{o}tes field.   This
sample includes $\sim1500$ AGN identified by the Stern wedge, with
spectroscopic or
photometric redshifts between $0.7<z<3$.  The shallow X-ray coverage of
this field results in few X-ray detections for the infrared-selected
AGN, but X-ray stacking shows that the infrared-selected candidates
are, on average, consistent with AGN activity.
These results show that while infrared and X-ray selected samples
do not identify exactly the same AGN populations, by and large they are
both effective at constructing samples dominated by AGN.

We note that broad consistency is found among these various works
even though a variety of survey data and selection techniques were
employed.  In particular, the different applications of mid-infrared
AGN color selection produce similar results, albeit with trade-offs
in completeness and contamination.  For comparison, we re-analyzed
our sample using the \citet{Lacy:04}\ wedge selection technique,
and we analyzed the \citet{Tajer:07}\ sample using the \citet{Stern:05b}\
wedge.  We find that the general trends and final conclusions are
unchanged, though the numbers and percentages do vary.  Studies,
such as that recently undertaken by \citet{Assef:09}, generally
find that the \citet{Lacy:04}\ selection technique is more complete
but is heavily contaminated by low-redshift star-forming galaxies.
In comparison, the \citet{Stern:05b}\ selection technique is less
contaminated by low- and intermediate-redshift galaxies but suffers
from some incompleteness, notably at $z \simeq 4.5$ as H$\alpha$
shifts through the IRAC 3.6~$\mu$m channel and for sources where the
AGN emission is dominated by host galaxy stellar emission. The
power-law AGN selection \citep{Alonso-Herrero:06, Donley:07} is the
least contaminated by non-AGN, but only selects a subset of the
\citet{Stern:05b}\ and \citet{Lacy:04}\ wedge-selected sources
\citep[see, e.g.,][]{Donley:08}.

\subsubsection{High-Redshift, Type-2 Quasar Selection}
\label{sexsiconclusion_sec:type2quasar}

\citet{Martinez-Sansigre:05}\ use the {\em Spitzer} FLS to search
for type~2 quasars at $z>2$ using the following selection criteria:
$f_{\rm 24 \mu m} \geq 300~\mu$Jy, $f_{\rm 3.6 \mu m} \leq 45~\mu$Jy,
and $350~\mu{\rm Jy} \leq f_{\rm 1.4 GHz} \leq~2$~mJy.  The $24~\mu$m
cut is designed to define a reliable catalog of luminous quasar
candidates, while the $3.6~\mu$m cut eliminates the unobscured
quasars and the radio cut eliminates starburst galaxies.  This
effort identified 21 high-redshift, type~2 quasar candidates, ten
of which were spectroscopically confirmed to be narrow-lined AGN at
$1.4<z<4$.  The other eleven sources exhibit featureless optical
spectra.

\begin{figure*}
\centering
\includegraphics[width=7in]{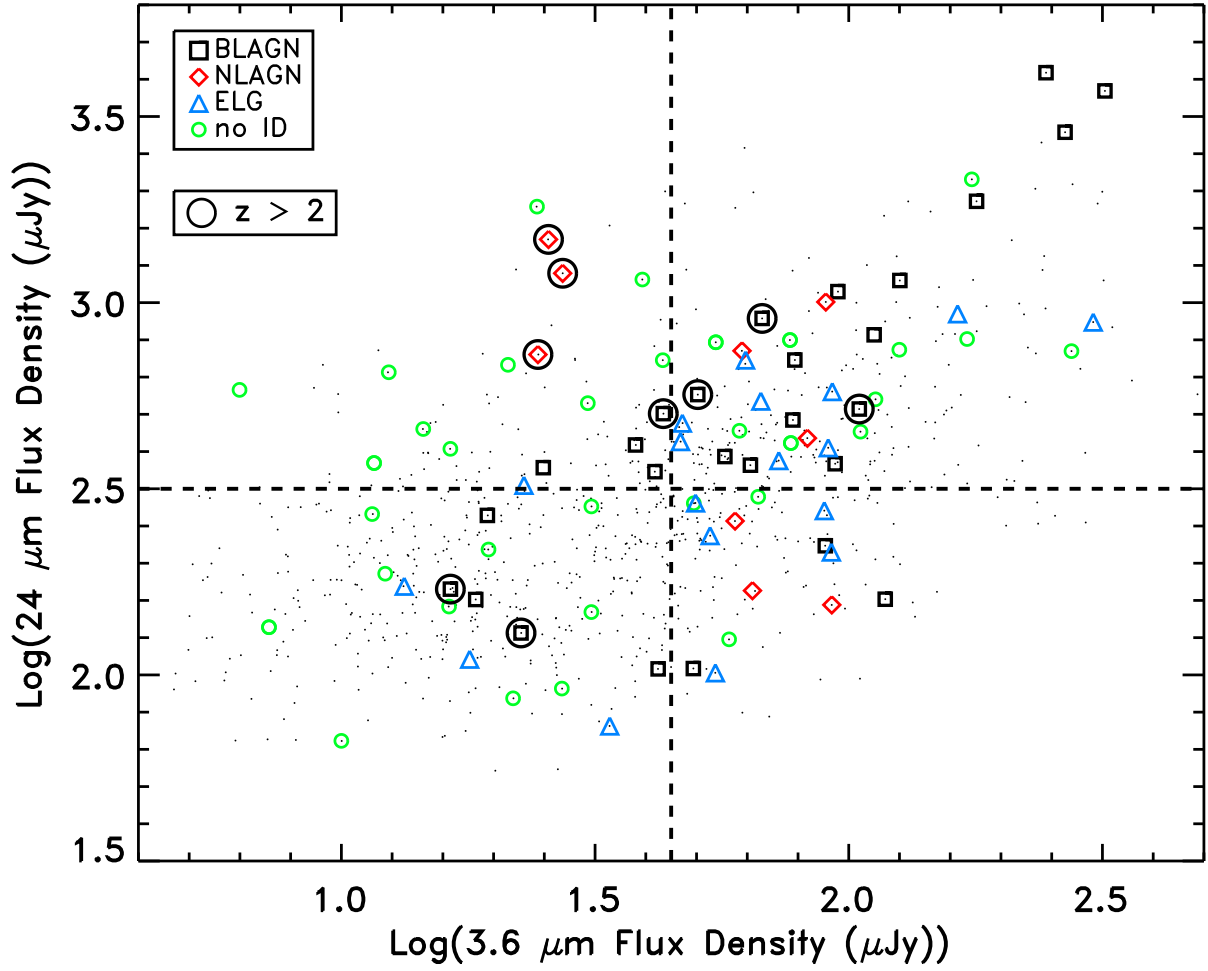}

\caption{The upper left quadrant of this plot indicates the infrared
selection area for highly obscured, $z>2$ type~2 AGN, defined by
\citet{Martinez-Sansigre:05}.  We plot sources with $\ge 5\sigma$ detections
at $3.6~\mu$m and $24~\mu$m.
Black circles surround all
spectroscopically confirmed $z>2$ sources.
This selection quadrant successfully identifies all three SEXSI-confirmed $z>2$ narrow-lined quasars. The ten spectroscopically unidentified
SEXSI sources (no ID) in the selection quadrant are faint in the optical 
(typically $R\simgt 24$) 
and have X-ray properties
that are consistent with the notion that they are indeed high-redshift, 
obscured quasars.}

\label{sexsiconclusion_fig:mips24_36}
\end{figure*}

Figure \ref{sexsiconclusion_fig:mips24_36}\ presents our sample of
sources detected at both $24~\mu$m and $3.6~\mu$m, with the upper-left
quadrant containing the sources that meet the \citet{Martinez-Sansigre:05}
infrared selection criteria.  The large symbols indicate SEXSI
\hardrange\ sources, with confirmed $z>2$ sources circled in black.
While our lack of deep radio data precludes application of the radio
cut, SEXSI X-ray detections provide another effective method to
avoid starburst contamination.  Of the subsample of SEXSI sources
with MIPS and IRAC coverage, we find that all three of the $z>2$
confirmed SEXSI NLAGN fall in the selection quadrant, while the
few confirmed broad-lined AGN selected with these criteria all 
fall along the edges of the selection box.  The high-redshift, narrow-lined sources 
exhibit the high-ionization UV emission lines indicative
of an active nucleus, and have hard ($2-10$~keV) X-ray luminosities
of $L_x \sim 10^{44.5}\lumin$.  These sources exhibit hard X-ray spectra, 
with best-fit \nh\ estimates of $1.4-9.4 \times 10^{23}$~cm$^{-2}$; 
two of the three sources do not have significant detections in the \softrange\ 
band. The sources at $z>2$ that exhibit
broad AGN lines fall outside the quadrant (one sits just at the edge).

The  
majority (10 of 17) of the X-ray-selected sources that fall in
the selection quadrant lack optical spectra. The SEXSI optical
followup program focused on sources with $R \simlt 23-24$, and thus
many of the fainter $3.6~\mu$m sources, which tend to also be faint
in $R$, lack spectroscopic coverage.  However, the combined X-ray
and optical properties of the unidentified sources are consistent
with high-luminosity, high-redshift, obscured AGN.  These ten sources
are not detected in our $R$-band imaging or have magnitudes of
$R>24$ and, if we assume $z=2$, have high X-ray luminosities ($L_{\rm
2-10~keV}\simgt 10^{44}~\lumin$).  The sources are hard in the
X-ray:  many do not have significant detections in the \softrange\ band. 
Using the median hardness ratio of the sample and assuming
$z=2$ and an intrinsic spectral 
index of $\Gamma=1.9$ we find \nh$\sim 3 \times 10^{23}$~cm$^{-2}$.  These properties
are similar to those exhibited by the $z>2$ NLAGN found by \citet{Martinez-Sansigre:05}, 
as well as by the spectroscopically identified $z>2$ NLAGN in our sample.

We thus confirm the result of \citet{Martinez-Sansigre:05}\ 
in a \hardrange\ X-ray-selected sample by 
showing that the $3.6~\mu$m$-24~\mu$m selection quadrant does indeed 
select high-redshift, highly obscured narrow-lined AGN. Instead of using 
1.4~GHz radio selection criterion to eliminate starburst contamination we 
rely on a \hardrange\ X-ray detection as an effective method to ensure
that the sources are active galaxies. All of the spectroscopically 
confirmed $z>2$ narrow-lined AGN in the SEXSI sample fall in the selection 
quadrant, and the SEXSI sources in the selection quadrant that lack spectroscopic 
information have X-ray hardness ratios and $R$-band characteristics consistent with 
the notion that they too are high-redshift, highly obscured quasars.

\section{X-ray Properties of Mid-Infrared-Selected AGN}
\label{sexsiconclusion_sec:stacking}

A large number of sources in our mid-infrared sample have four-band IRAC detections but 
lack significant {\em Chandra} detections. In this section we present results from an X-ray stacking 
analysis to study the mean X-ray properties of these sources, in particular 
focusing on the subset of these sources that fall within the mid-infrared
AGN selection wedge. The aim is to first understand whether the mean 
X-ray properties of the wedge-selected sources differ from those outside
the wedge, and whether these differences are consistent with the idea 
that the wedge-selected sources are active galaxies. A second goal is 
to estimate the intrinsic obscuration of the wedge-selected 
population using the hardness ratio of the stacked X-ray signals and assuming
various population redshifts.   
We first describe the stacking methodology, and then 
present the results (\S \ref{sexsiconclusion_sec:stackingresults}) and
a discussion (\S \ref{sexsiconclusion_sec:discussion}).

\subsection{X-ray Stacking Methodology}
\label{sexsiconclusion_sec:stackingmethod}

The X-ray stacking analysis used tools provided by CIAO and was
aided greatly by {\tt acis\_extract}\footnote{Available at
http://www.astro.psu.edu/xray/docs/TARA/ae\_users\_guide.html.}
\citep[ver.~3.91;][]{Broos:02}.  We use CIAO (ver.~3.2) and CALDB
(ver.~3.1) for the analysis presented here.  Source lists for X-ray
spectral extraction were assembled using our SExtractor catalogs of
four-band IRAC sources (with $>5 \sigma$ detections in each band)
that fall within the {\em Chandra} images,
excluding sources very near the edge of an ACIS chip.\footnote{These
mid-infrared positions were then `uncorrected': IRAC $3.6~\mu$m
source positions were shifted to the uncorrected SEXSI X-ray frame
by adding to each mid-infrared position the average X-ray-to-optical
offset correction \citep[see \S 3 and Table 2 of][]{Eckart:05}.}
Sources that fall within 1~Mpc of the center of each galaxy cluster
target are eliminated from the stacking to reduce contamination from cluster galaxies.

Spectra were extracted for each source in the resulting catalog
($\sim 2500$ sources).  We choose 1.5 keV as the energy
at which the PSFs are computed and use extraction apertures 
that enclose 80\% of the energy (0.8$\times$ PSF). 
We calculated individual auxiliary response files (ARFs) and redistribution matrix
files (RMFs) for each source, and we extracted a
background spectrum from a local circular background
region that includes at least 100 counts, taking care to mask out
all detected X-ray sources (SEXSI sources, including soft-only X-ray
sources, target point sources, and extended cluster emission).  We scaled the
background spectra based on the ratio of total exposure
in the source extraction region to that of the background region.

Source counts and scaled background counts were tabulated for each
source in several energy bands: the standard $0.5 - 8$ keV (full
band), $2 - 8$ keV (hard band), and $0.5 - 2$ keV (soft band), as
well as $2-4$ keV, $4-6$ keV, and $6-8$ keV.  Estimates of individual
source significance were calculated for each band. Following
\citet{Laird:05}, we define detection significance as $S/\sqrt{B}$
and signal-to-noise ratio as $S/\sqrt{S+B}$, where $S$ and $B$ are
the net source counts and scaled background counts, respectively.
The average exposure per pixel (in each energy band) was calculated
for each source by averaging the ARF over the given energy band and
then over the source extraction cell.  In addition to extracting
X-ray spectra at the position of the IRAC sources, as a control 
sample we repeated this procedure using a catalog with a 1\arcmin\ offset 
applied to all source positions.

After calculating individual source statistics we  stacked
the sources.  First, any IRAC sources that are within 5\arcsec\ of
an X-ray source were eliminated from the sample.  In addition, we
eliminated IRAC sources that lie within 5\arcsec\ of 64 sub-threshold
X-ray sources that were initially identified by {\tt wavdetect} but
later eliminated from the main SEXSI catalogs because they fall
below the strict SEXSI $P<10^{-6}$ significance threshold.\footnote{In
\citet{Harrison:03}\ we used {\tt wavdetect} for initial source
identification with the {\tt sigthresh} parameter set to $10^{-5}$
to produce a complete catalog. In a subsequent step we tested the
significance of each source individually and eliminated sources
from the source catalog with a nominal chance occurrence probability
greater than $10^{-6}$.  Sources that fall between the two thresholds
were not included in the SEXSI X-ray sample.} In field CL0848+44
we include only IRAC sources that fall on ACIS chip~3, since
the majority of the other X-ray sources are at very large off-axis
angles.

The total signal in each band was calculated by summing the net
background-subtracted counts from each source in the sample. 
To calculate an average flux we divided the summed signal
by the sum of the average exposure per pixel in each extraction
cell to convert from counts to ph cm$^{-2}$ s$^{-1}$.  To convert
from ph cm$^{-2}$ s$^{-1}$ to $\fluxu$, we assumed a power-law
spectrum with photon index $\Gamma=1.5$.

\subsection{Stacking Results}
\label{sexsiconclusion_sec:stackingresults}

We stacked three basic samples: all X-ray undetected sources (1298
sources), the wedge-selected sources (226 sources), and those that
fall outside of the wedge (1072 sources).
Table~\ref{sexsiconclusion_tbl:stackresults}\ presents a summary
of the stacking results, where $N$ is the number of 
sources in the particular sample. The ``Cts'' column gives the total 
number of background-subtracted
net counts in the sample. The ``Sig'' column gives 
an estimate of the significance of the stacked counts
($S_{\rm tot}/\sqrt{B_{\rm tot}}$, see \S \ref{sexsiconclusion_sec:stackingmethod}) 
and $f_x$ is the average X-ray flux
for that band in units of $10^{-17} \fluxu$.  
The full sample of all X-ray undetected,
four-band IRAC sources showed a significant detection in the stacked
{\em Chandra} images.
The stacks of the sources created
by the shifted source catalog showed no significant detection
in any band.

Stacking only the wedge-selected sources, we find significant
detections in all three X-ray bands.  
The wedge-selected sample is, on average, brighter in the X-ray:
the average $0.5-8$~keV X-ray flux from the wedge-selected IRAC sample 
is $\sim 2$ times larger than that of the full sample of four-band
IRAC sources.
This trend is most
significant in the hard band, where the average flux of the
wedge-selected sample is over three times that of the sample that
lies outside the wedge.  These findings are consistent with the
idea that AGN preferentially lie inside the wedge.

To allow for errors in the IRAC photometry (e.g., see Figure~\ref{sexsiconclusion_fig:wedgeplot}), 
we also stacked based
on an extended wedge defined by:
\begin{equation}
\begin{split}
([5.8]-[8.0]) > 0.5  \\
 \wedge  ~~([3.6]-[4.5])>0.2\cdot ([5.8]-[8.0])+0.078   \\
  \wedge ~~ ([3.6]-[4.5])>2.5 \cdot([5.8]-[8.0]) -3.77~, 
\end{split}
\end{equation}
\noindent where $\wedge$ is the logical AND operator. This extended
wedge is constructed by taking the original \citet{Stern:05b}\ wedge
and adding a swath of width 0.1 mag at each edge.  The stacking results
for the samples inside
and outside the extended wedge are also presented in
Table~\ref{sexsiconclusion_tbl:stackresults}; the extended wedge
swath includes an additional 150 sources. Removing these 150 sources
from the 1072 sources not selected by the original wedge removes
a third of the stacked counts in the hard-band, leaving a source
significance of only 3.9 in the hard-band sample of sources outside
of the extended wedge.  This exercise demonstrates that the hard
X-ray emission, which is more easily produced by AGN than by star
formation, comes primarily from sources in or just outside the
wedge, at least at our detection sensitivity.

Comparing the stacked fluxes of wedge-selected sources to those outside the
wedge we find that the mid-infrared-selected AGN are about three
times brighter in the hard band than those outside the wedge, which is the same ratio
we find if we calculate the average flux for SEXSI \hardrange
-detected sources inside and outside the wedge.  Although the
wedge-selected sources are harder than those outside the wedge, the
column density implied by the average hardness ratio is not extreme.
Table~\ref{sexsiconclusion_tbl:stacknh}\ provides estimated \nh\
values based on the hardness ratio of the stacked spectrum of the wedge-selected sample for
assumed population redshifts ranging from 0.0 to 3.0. The \nh\
values are calculated assuming an underlying power-law spectrum
with $\Gamma=1.9$ and a Galactic column density of $2 \times 10^{20}$
cm$^{-2}$; the associated uncertainties are estimated by propagating the Poisson
sampling error in the hardness ratio.
These \nh\ estimates range from $\sim 1.3 \times 10^{22}$
to $4.8 \times 10^{23}$~cm$^{-2}$, suggesting that the sources are
intrinsically obscured, but are not, on average, Compton thick.

\subsection{Discussion of Stacking Results}
\label{sexsiconclusion_sec:discussion}

The hardness ratio of the wedge-selected stacked spectrum is consistent with
moderate intrinsic obscuration, but is not suggestive of a highly
obscured, Compton-thick source population.  It is possible that the
stacking procedure is missing a population of sources, such that the stacked
counts are dominated by a subset of the less obscured wedge-selected
sources.\footnote{We estimate that a lower limit of $\approx 30$\% of the sources 
must be contributing to the stacked signal: if it were fewer than 30\% then 
individual source fluxes would fall above the {\em Chandra} detection threshold 
in order to reproduce the average stacked flux in the $2-8$~keV band. This provides
an upper limit
of 70\% of the sources that are not
contributing significantly to the stacked signal.}
Another possibility is that the assumption of an
intrinsic $\Gamma=1.9$ power-law component plus photoelectric
absorption is too simple, so that the estimated \nh\ values are not
representative of the sample.  For example, it is possible that the 
sources within the sample have a range 
of power-law indices and/or absorbing columns. 
In addition, the soft X-ray emission may originate
in a different location than the hard, power-law component, skewing
the \nh\ estimates to lower values.  In reality, it is likely that some
combination of these and other effects is operating.

\citet{Hickox:07}\ use an IRAC color-selection technique to identify
a large sample of AGN from the IRAC Shallow Survey, and an
infrared/optical selection technique to separate the sources into
unobscured (IRAGN1) and obscured (IRAGN2) sub-samples.  They perform
a similar stacking analysis of complementary X-ray data from the
shallow (5~ks) {\em Chandra} XBo\"{o}tes survey and find that the
IRAGN1 are consistent with no obscuration and the IRAGN2 are
consistent with \nh$\sim 3 \times 10^{22}$~cm$^{-2}$, independent of 
photometric redshift from $z_{\rm phot}\approx 0.7-2.5$.  The implied
obscuring column density found using this shallow, wide-area data
set is consistent with that of the SEXSI stacking analysis assuming 
$\langle z_{\rm stack} \rangle \simlt 1$. If the majority of the 
stacked emission in the SEXSI wedge-selected sample 
comes from a higher redshift population, then the implied
\nh\ is higher than that found in the XBo\"{o}tes analysis.  The
average X-ray fluxes of X-ray undetected infrared AGN
from \citet{Hickox:07}\ are $f_{\rm 2-7~keV} \sim 2 \times 10^{-15}\fluxu$,
which is just at the hard-band SEXSI X-ray detection threshold, and more than an order of
magnitude larger than the average stacked hard-band fluxes from the SEXSI analysis.

Several groups have explored the X-ray properties of small samples
of optically identified type~2 quasars from the Sloan Digital Sky
Survey using {\em Chandra} and {\em XMM-Newton} 
\citep[e.g.,][]{Vignali:06,Ptak:06}. These studies have found
that approximately half of the type~2 quasars are detected in the
$\approx 5-40$~ks X-ray images, and those that are detected 
exhibit moderate obscuring column
densities ($\sim 10^{22}-10^{23}$~cm$^{-2}$).  The sources that are
undetected in the X-ray are likely highly obscured such that
little X-ray radiation below 10~keV escapes.  The high bolometric
luminosities of these sources, as evidenced by both their mid-infrared
properties and their optical spectra, imply highly obscured, 
high-luminosity AGN that will require a sensitive high-energy ($\simgt
10$~keV) X-ray mission before they are detected in the X-rays.

\section{Summary and Conclusions}
\label{sexsiconclusion_sec:conclusions}

We have presented an exploration of the relative 
efficiency of selecting AGN
in the X-ray with {\em Chandra} and in the mid-infrared with {\em
Spitzer}.  In this section we summarize our comparison of the
selection techniques and make predictions about the AGN samples that will
be selected with future X-ray and mid-infrared surveys.

\subsection{Summary}
\label{sexsiconclusion_sec:midIRsummary}

Two-thirds of the X-ray-selected AGN are also selected
by the \citet{Stern:05b}\ {\em Spitzer} AGN selection criteria.
Nearly all of the SEXSI X-ray sources that are classified as either broad-line
or narrow-line AGN (i.e., those that exhibit high-ionization optical/UV
emission lines in their optical spectra) are consistent with wedge
selection within photometric uncertainties.

A large fraction of the low luminosity ($L_{\rm 2-10~keV} < 10^{43.5} \lumin$), 
hard X-ray (\hardrange)
AGN identified by {\em Chandra} will be missed by {\em Spitzer} selection techniques.  
These low X-ray luminosity sources tend to lack the high-ionization emission lines
that allow AGN identification via optical spectroscopy; thus, the
sources that are missed by the {\em Spitzer} color-color AGN selection
will also remain unidentified in optical spectroscopic surveys.
These sources tend to lie at lower redshift and have {\em higher}
fluxes in the infrared, on average.  
This suggests that many of these non-wedge-selected X-ray sources
have mid-infrared fluxes that are dominated by stellar (starburst) emission as
opposed to AGN activity;
indeed, most have optical
spectra of ELGs, consistent with the conclusion that emission related to star
formation outshines the emission from the active nucleus in some 
parts of the spectrum (i.e., in the mid-infrared and optical), while
the X-ray luminosities of these sources are dominated by X-ray emission from the active 
nuclei.
In addition, we find that while a fair fraction ($\sim 63$\%) of
the optically bright ($R<21$) wedge-selected {\em Spitzer} AGN are
also identified by {\em Chandra}, over 80 percent of the optically faint
wedge sources are not detected in the X-ray.  It is likely that these fainter
sources are AGN that are heavily obscured and/or lie at high redshift.
We also find that X-ray-selected high-redshift, type-2 quasars
are also selected via the $3.6~\mu$m$-24~\mu$m 
selection criteria proposed by \citet{Martinez-Sansigre:05}; the
radio selection criteria used by those authors 
to eliminate starburst contaminants is not necessary when the
sources have X-ray properties indicative of AGN activity.

To explore the mean X-ray properties of the X-ray undetected,
IRAC-selected AGN candidates, we have stacked the  {\em
Chandra} data extracted from the positions of IRAC sources. 
The stacked wedge sources show significant X-ray
signals in the full, soft, and hard X-ray bands. The average hardness
ratio of the stacked spectrum of sources inside the wedge is higher than that
of sources outside of the wedge, and the average $2-8$~keV flux is three times
larger.  The hardness ratio of the wedge-selected stacked spectrum is consistent
with moderate intrinsic obscuration (\nh$\sim
10^{22}-10^{23.7}$~cm$^{-2}$), but is not suggestive of a highly
obscured, Compton-thick source population.

\subsection{Conclusions}
\label{sexsiconclusion_sec:finalconclusions}

The results of our study illustrate the challenges of identifying obscured and low-luminosity
AGN  using any one technique.  A complete understanding of AGN activity and black hole
growth will require multiwavelength data sets: low-luminosity AGN in galaxies with active
star formation cannot be selected using mid-infrared color techniques, but are effectively
found at low redshift with the current-generation of X-ray telescopes;  obscured
high-redshift objects are beyond the sensitivity threshold of {\em Chandra} and {\em XMM},
but they are effectively identified in the mid-infrared if the AGN luminosity is at the high end of
the luminosity distribution.    

\begin{figure}
\centering
\includegraphics[width=6in]{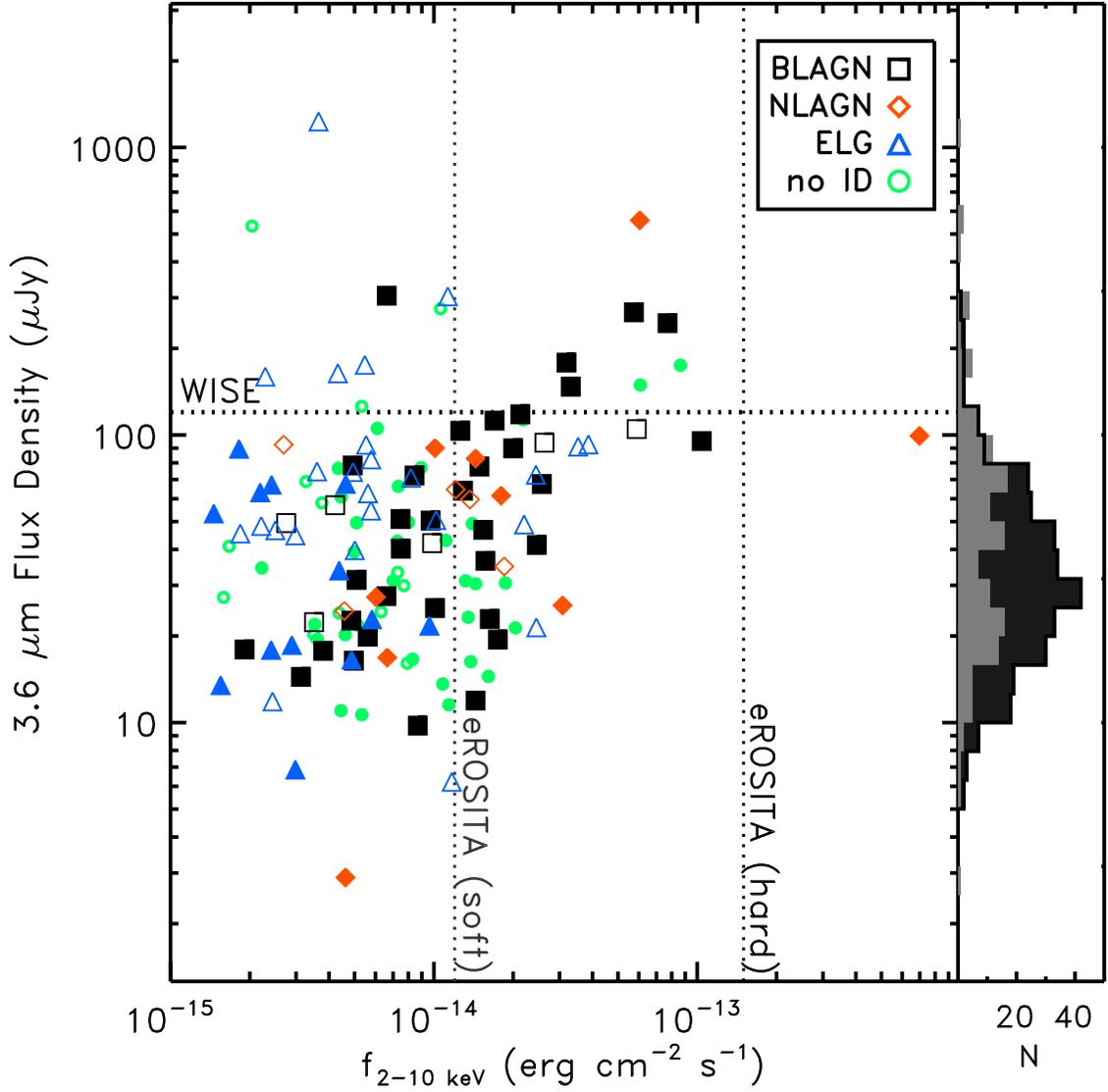}
\caption{Mid-infrared flux density vs. hard X-ray flux for SEXSI
sources.  
Filled symbols indicate sources within the Stern wedge. The histogram
on the right shows the $3.6~\mu$m flux distribution for SEXSI AGN
(light gray) and all IRAC wedge sources from the SEXSI {\em Spitzer}
fields (black).  The dotted lines illustrate the depths that will
be reached by two upcoming full-sky survey satellites:  the {\em
Wide-field Infrared Survey Explorer} ({\em WISE}; horizontal line)
and {\em eROSITA} (vertical lines).  The line labeled ``eROSITA (hard)'' provides
the expected \hardrange\ sensitivity while the ``eROSITA (soft)'' line gives an indication
of the \softrange\ sensitivity, scaled to the \hardrange -band using $\Gamma=1.9$. 
The {\em eROSITA} effective area is much larger below 2~keV, thus the mission will 
detect many more AGN in the soft band. Some SEXSI AGN brighter than the ``eROSITA (soft)'' 
limit but with spectra harder than $\Gamma=1.9$ will in fact not be detected by {\em eROSITA}.}
\label{sexsiconclusion_fig:f36vsfx}
\end{figure}

The results from current X-ray and mid-infrared satellites enable us to 
estimate the progress possible
with upcoming space missions that plan to conduct all-sky or pencil beam surveys.
The {\em Wide-field Infrared Survey Explorer} ({\em WISE;} \citealt{Mainzer:05}), scheduled for launch
at the end of 2009, will provide full-sky mid-infrared imaging
reaching a depth of 120~$\mu$Jy at 3.3~$\mu$m (similar to IRAC
channel~1) and 160~$\mu$Jy at 4.7~$\mu$m (similar to IRAC channel~2).
At these shallow depths, almost every source redder than [3.3]$-$[4.7]
$\approx 0.5$ will be an AGN; high-redshift galaxies are too faint
to make this flux cut, and the coolest brown dwarfs are too rare.
The primary contaminant will be actively star-forming galaxies at
redshifts of a few tenths, which should be readily identifiable
from their optical morphologies and colors in relatively shallow
imaging such as is available from the Sloan Digital Sky Survey.  
By applying a flux cut at the {\em WISE} depths to 
the Bo\"{o}tes IRAC survey \citep{Eisenhardt:04,Ashby:09}, 
we predict that {\em WISE} should
detect approximately 85 AGN candidates per square degree, of which
$\approx 15\%$ will be low-redshift star-forming galaxy contaminants.
{\em WISE} will therefore provide an unprecedented sample of high-luminosity 
obscured AGN across a range of redshifts up to  $z \sim 3$ identified based
on mid-infrared color selection alone.    {\em WISE} will also detect numerous other
low-redshift ($z < 1$), low-luminosity ($L_{\rm 2-10~keV} \simlt 10^{43} \lumin$) active galaxies, 
but the significant number of open
symbols above the {\em WISE} threshold shown in Figure~\ref{sexsiconclusion_fig:f36vsfx}
shows that mid-infrared colors alone are insufficient to identify them as AGN.

The {\em extended R\"ontgen Survey with an Imaging Telescope Array} ({\em eROSITA;} \citealt{Predehl:06})
X-ray mission, planned for launch in 2011,  will provide full-sky hard X-ray images
to a depth of $f_{\rm 2-10~keV} = 1.5 \times 10^{-13}~\fluxu$.  
Using the sample of XBo\"{o}tes sources \citep{Kenter:05} with fluxes above the {\em eROSITA} 
hard-band sensitivity limit, 
we predict that {\em eROSITA} should detect $\approx 4$ hard X-ray sources per square degree.
The mission will detect many more AGN in the soft X-ray band because
the effective area is much larger at energies below 2~keV, reaching soft-band
depths of $f_{\rm 0.5-2~keV} = 9 \times 10^{-15}~\fluxu$.  Scaling from the 
XBo\"{o}tes \softrange\ sample we predict that {\em eROSITA} will detect
$\approx 120$ sources per square degree in the soft band.
Figure~\ref{sexsiconclusion_fig:f36vsfx} shows that {\em eROSITA} will identify a large population of
AGN below the {\em WISE} threshold, most of which we project will be at $z \simgt 1$.  
The majority of these sources will be unobscured 
BLAGN, but a significant population will be narrow-line, obscured AGN (NLAGN and ELG).    
However, {\em eROSITA} does
not reach sufficient depth to detect many of the low-luminosity, low-redshift AGN detected, but not
identified as such, by {\em WISE}.

Clearly {\em WISE} and {\em eROSITA} combined will miss (or mis-identify) low-luminosity obscured 
AGN at all redshifts, most of which will also not be found in optical and soft X-ray surveys.   
At low redshift only a sensitive X-ray mission extending to E$_x > 10$~keV will 
effectively uncover this population, which will be visible only in reprocessed light, 
if at all, at E$_x < 10$~keV. The  {\em Nuclear Spectroscopic Telescope Array}
({\em NuSTAR;} \citealt{Harrison:05})\footnote{\tt http://www.nustar.caltech.edu.}, scheduled
for launch in 2011,  
will make major advances in studying faint extragalactic 
sources at energies above 10~keV. 
{\em NuSTAR} will be the
first focusing, high-energy X-ray satellite and is designed to 
make targeted observations of the hard X-ray sky from $6-79$~keV.  
{\em NuSTAR}, which has a field-of-view comparable to {\em Chandra},
plans to cover  the Bo\"{o}tes and Great Observatories
Origins Deep Survey (GOODS) fields, and should resolve $\simgt 50$\% of the 
30~keV X-ray background, 
reaching $10-40$~keV flux limits of $< 10^{-14}~\fluxu$. 
Many of the
resolved sources are expected to be low-luminosity, nearby AGN not identified as such
in current mid-infrared or \hardrange\ X-ray surveys, 
and some {\em NuSTAR} sources may be AGN
with hard spectra owing to low-efficiency accretion.
For studying the high-redshift, low-luminosity population, the {\em International X-ray Observatory} ({\em IXO}) 
will reach \hardrange\ 
flux limits of $\sim 5 \times 10^{-17}~\fluxu$ in a megasecond exposure with its  wide-field imager.
It could survey up to 10~deg$^2$ to a depth of
$10^{-16}~\fluxu$ and 0.25~deg$^2$ to a depth of $10^{-17}~\fluxu$, uncovering the largest 
population of high-redshift AGN yet detected.

Predictions for the AGN populations that will be uncovered by these future missions
must rely on careful studies of the AGN identified by existing X-ray and
infrared surveys. As illustrated in this paper, an understanding of the 
AGN selection techniques of the various missions, and their relative strengths 
and weaknesses, is crucial as datasets are synthesized to obtain
an understanding of black hole growth and evolution in the 
universe.

Finally, we attempt to put the results of our study in the context of understanding
the cosmic history of black hole growth and AGN accretion modes.
The primary conclusions of this paper are that heavily obscured
luminous AGN are often missed by X-ray selection, while low-luminosity
AGN are often missed by mid-infrared selection.  For the high-luminosity
sources, various work has shown that, at least for the optically-selected
unobscured sources, most are emitting at close to their Eddington
limit \citep[e.g.,][]{Kollmeier:06}.  Multiwavelength observations
are consistent with the heavily-obscured high-luminosity sources
having similar intrinsic SEDs, simply with their UV, optical, and
X-ray emission suffering from absorption.

As for the low-luminosity AGN, there are two possibilities:  they
could be lower mass black holes again emitting at close to their
Eddington limits, or they could be average-sized black holes, with
masses similar to the high-luminosity sources, but simply emitting
at lower Eddington ratios.  Using a new set of empirical, low-resolution
SED templates for AGN and galaxies, \citet{Assef:09} shows that the
likelihood of a source to be selected as an AGN using the
\citet{Stern:05b} mid-infrared color criteria is primarily a function of
how strong the AGN is relative to the host galaxy --- e.g., the
Eddington ratio assuming that, to first order, the host galaxy
mid-infrared emission scales with the galaxy total 
stellar mass which scales with the nuclear
supermassive black hole mass.  Thus, the fact that we find few
low-luminosity sources using the mid-infrared criteria suggests that the
low-luminosity sources are {\em not} emitting at close to their
Eddington ratio.  This is consistent with the results of \citet{Babic:07}\
who find that low-luminosity X-ray sources in the {\em Chandra}
Deep Field-South have a wide range of Eddington ratios, $10^{-5}
\simlt \log(L_{\rm bol}/L_{\rm Edd}) \simlt 1$.  One caveat on
this interpretation is that it assumes that the AGN SED is independent
of both Eddington ratio and luminosity --- the former of which has
recently been questioned by \citet{Vasudevan:07}.  Future work, in
particular, higher energy observations with {\em NuSTAR}, will
quantify the extent to which the intrinsic AGN SEDs depend on both
luminosity and Eddington ratio.

\acknowledgements
This work is based on observations made with the {\it Spitzer Space
Telescope}, which is operated by the Jet Propulsion Laboratory,
California Institute of the Technology under contract with the
National Aeronautics and Space Administration (NASA).  Support for
this work was provided by NASA through award number 1314516 issued
by JPL/Caltech.  The authors thank the anonymous referee for his/her
careful read and unusually diligent comments as well as
Lewis Kotredes for assistance
with {\em Spitzer} data reduction.  MEE acknowledges support from
the NASA Postdoctoral Program.

\facility{{\it Facilities:} \facility{Spitzer}, \facility{CXO}, \facility{Keck:I}, and \facility{Keck:II}.}

\clearpage 
\bibliographystyle{apj}

\clearpage

\begin{deluxetable*}{lcrccl}
\tabletypesize{\footnotesize}
\tablecaption{{\em Spitzer} observations of SEXSI fields.}
\tablecolumns{6}
\tablewidth{0pc}
\tablehead{ & & \colhead{{\em Chandra}} & \colhead{{\em Spitzer}} & \colhead{} & \\
Cluster Field & \colhead{$z$} & \colhead{Exp. Time\tablenotemark{a}} & \colhead{Instrument} & \colhead{PID\tablenotemark{b}} & notes}
\startdata
CL0848+44 & 1.27  & 186~ks & IRAC & 00017 & \\
&  &  & IRAC & 00064 & \\
&  &  & IRAC & 20694 & wider FOV \\
&  &  & MIPS & 00083 &  \\
&  &  & MIPS & 20694 &  wider FOV \\
RXJ0910+54 & 1.11& 171~ks &  IRAC &  00017  & \\
& & & IRAC &  {\bf 20808}  & wider FOV\\
& & & MIPS &  00083  & \\
& & & MIPS &  {\bf 20808}  & wider FOV\\
RXJ1317+29 & 0.81 & 111~ks &  IRAC &  00017  & \\
BD1338+29  & 0.64 & 38~ks &  IRAC &  {\bf 20808}  & \\
 &  & & MIPS &  {\bf 20808}  & \\
RXJ1716+67 & 0.81  & 52~ks &IRAC &  00017  & \\
&  & & IRAC &  {\bf 20808}  & wider FOV\\
&  & & MIPS &  {\bf 20808}  & \\
RXJ2247+03 &0.18&  49~ks & IRAC &  {\bf 20808}  & \\
& & & MIPS &  {\bf 20808}  & \\
\enddata
\tablenotetext{a}{This column provides the {\em Chandra} exposure times 
after removal of background flares.}
\tablenotetext{b}{{\em Spitzer} program identifier (PID) 00017 ---
distant X-ray galaxy clusters (P.I. Fazio; GTO); PID 00064 ---
combined program (P.I. Fazio; GTO); PID 00083 --- use of massive
clusters as cosmological lenses (P.I. Rieke; GTO); PID 20694 ---
IRAC and MIPS mapping of galaxy populations in a supercluster at
$z=1.27$ (P.I. Stanford; GO2); {\bf PID 20808} --- SEXSI (P.I.
Stern; GO2).}

\label{sexsiconclusion_tbl:data}
\end{deluxetable*}

\clearpage
\begin{landscape}
\begin{deluxetable}{lrrrrrrrrrrrrrrrrrrrrrrrr}
\tablecaption{SEXSI-{\em Spitzer} Catalog.}
\pagestyle{empty}
\tabletypesize{\tiny}
\tablecolumns{25}
\tablewidth{0pc}
\tablehead{ 
CXOSEXSI &
\colhead{$\Delta \alpha$} &
\colhead{$\Delta \delta$} &
\colhead{$S_{\rm 2-10}$} &
\colhead{$HR$} &
\colhead{$R$} &
\colhead{class} &
\colhead{$z$} &
\colhead{log($L_x$)} &
\colhead{log(\nh)} &
\colhead{$f_{3.6}$} &
\colhead{$f_{4.5}$} &
\colhead{$f_{5.8}$} &
\colhead{$f_{8.0}$} &
\colhead{$f_{24}$}\\
\colhead{} & \colhead{[\arcsec]} & \colhead{[\arcsec]} & 
\colhead{[$10^{-15}$]} & \colhead{} & \colhead{[mag]} &
\colhead{} & \colhead{} & \colhead{} & 
\colhead{} & \colhead{[$\mu$Jy]} & \colhead{[$\mu$Jy]} & 
\colhead{[$\mu$Jy]} & \colhead{[$\mu$Jy]} & \colhead{[$\mu$Jy]} \\ 
\colhead{(1)} & \colhead{(2)} & \colhead{(3)} & 
\colhead{(4)} & \colhead{(5)} & \colhead{(6)} &
\colhead{(7)} & \colhead{(8)} & \colhead{(9)} & 
\colhead{(10)} & \colhead{(11)} & \colhead{(12)} & 
\colhead{(13)} & \colhead{(14)} & \colhead{(15)} } 
\startdata
J084809.8+444901 & -0.7 &  0.1  &  35.50 &-0.26 &   20.73 &  unid  &       &      &           &   82.5$\pm$ 8.3&          &  133.9$\pm$ 13.6 &           &               \\
J084811.7+445302 & -1.7 &  0.1  &   8.01 & 0.44 &   23.46 &  unid  &       &      &           &   49.8$\pm$ 5.0&   48.6$\pm$ 4.9 &   48.0$\pm$  5.3 &   55.7$\pm$  6.1 &               \\
J084812.4+445657 &  0.3 &  0.5  &  15.40 &-0.32 &   19.45 &  unid  &       &      &           &         &  210.2$\pm$21.1 &           &  584.4$\pm$ 58.6 &               \\
J084818.4+444844 & -0.9 & -0.0  &   9.18 &-0.40 &   20.66 &   ELG  & 0.405 & 42.7 & $<$   &   65.7$\pm$ 6.6&          &   47.4$\pm$  4.9 &           &               \\
J084820.8+445648 &  1.8 & -1.5  &   4.55 &-0.15 &$>$24.40 &  unid  &       &      &           &         &          $<$8.5 &           &          $<$18.0 &               \\
J084822.2+445223 & -2.0 & -0.7  &   3.54 &-0.04 &   23.32 & BLAGN  & 2.187 & 43.9 &    20.8   &         &          &           &           &   $<$  76.0 \\
J084822.2+445627 & -1.5 & -1.2  &  10.90 &-0.26 &         & unid   &      &           &       &         &  346.1$\pm$34.7 &           &  173.1$\pm$ 17.9 &              \\
J084824.6+445355 & -0.8 &  0.3  &   2.61 & 0.03 &   22.21 &   ELG  & 0.747 & 42.7 &    22.4   &         &          &           &           &  711.8$\pm$ 101.1 \\
J084825.1+444808 & -1.7 & -0.2  &   7.61 &-0.55 &   20.73 & BLAGN  & 1.320 & 43.7 & $<$  &   84.5$\pm$ 8.5&          &  104.5$\pm$ 10.6 &           &             \\
J084827.2+445433 & -0.6 & -1.0  &   9.98 &-0.40 &   20.33 & BLAGN  & 0.899 & 43.5 & $<$ &         &          &           &           & 1228.5$\pm$ 155.5 \\
J084827.4+445604 &  2.1 & -1.6  &   4.56 & 0.31 &   23.67 &   ELG  & 1.528 & 43.7 &    22.7   &         &          &           &           &  564.4$\pm$  86.1 \\
J084830.2+445604 &  0.0 &  0.0  &   2.55 &-0.18 &$>$24.40 &  cont  &       &      &           &         &          &           &           &  $<$  65.9 \\
J084831.6+445442 & -0.3 &  0.7  &   2.44 &-0.15 &   25.42 &   ELG  & 1.267 & 43.2 &    22.0   &   11.8$\pm$ 1.2&   11.5$\pm$ 1.2 &   10.6$\pm$  1.6 &    8.5$\pm$  1.3 &  $<$  90.6 \\
J084832.7+445711 &  0.0 &  0.0  &   2.62 & 0.57 &   22.01 &   ELG  & 0.749 & 42.7 &    23.0   &         &                 &           &           &  $<$  87.1 \\
J084834.4+444937 &  1.2 &  0.0  &   3.02 & 0.15 &$>$24.40 &  unid  &       &      &           &    9.5$\pm$ 1.0&                 &   12.0$\pm$  1.7 &           &  $<$ 104.2 \\
\enddata
\tablecomments{Columns (4)--(10) present X-ray and
optical source information from previous SEXSI 
publications \citep{Harrison:03, Eckart:05, Eckart:06}.}
\label{sexsiconclusion_tbl:sexsispitzercat}
\end{deluxetable}
\clearpage
\end{landscape}

\begin{deluxetable*}{lccccc}
\tabletypesize{\footnotesize}
\tablecaption{Wedge selection as a function of X-ray luminosity.}
\tablecolumns{6}
\tablewidth{0pc}
\tablehead{\colhead{} & \colhead{}& \colhead{} & \multicolumn{3}{c}{\underline{~~~$L_{\rm min} < L_x < L_{\rm max}$~~~}} \\
Sample & log($L_{\rm min}$) & log($L_{\rm max}$) & \colhead{Wedge} & \colhead{Total} & \colhead{\%} }
\startdata
All classes   & --- & 43.0 &  5  & 14 & 36\% $\pm$ 16\%\\
BLAGN &  &  & 1 & 2 & 50\%  \\ 
NLAGN &  &  & 0 & 1 &  0\% \\ 
ELG &    &  & 4 & 11 &  36\%  \\ 
\hline
All classes & 43.0 & 43.5 & 13 & 26 & 50\% $\pm$ 14\% \\
BLAGN &   &  & 4 & 5 & 80\% \\ 
NLAGN &  & & 2 & 2&  100\% \\ 
ELG & & & 7 & 19 & 37\%  \\ 
\hline
All classes & 43.5 & 44.0 & 14 & 23 & 61\% $\pm$ 16\% \\
BLAGN & & & 11 & 14 & 79\% \\ 
NLAGN & & & 1 & 2 & 50\% \\
ELG   & & & 2 & 7 & 29\% \\ 
\hline
All classes & 44.0 & --- & 23 & 28 & 82\% $\pm$ 17\%\\
BLAGN & & & 17 & 18 &  94\% \\ 
NLAGN & & & 6 & 9 & 67\% \\
ELG & & & 0 & 1 &   0\% \\         
\enddata
\tablecomments{The percentage errors represent the Poisson sampling uncertainties.}
\label{sexsiconclusion_tbl:wedgelum}
\end{deluxetable*}

\begin{deluxetable*}{lcccccccccccc}
\tabletypesize{\footnotesize}
\tablecaption{Wedge selection as a function of optical magnitude.}
\tablecolumns{14}
\tablewidth{0pc}
\tablehead{\colhead{} & \multicolumn{6}{c}{\underline{Bright: $R<21$}} & \multicolumn{6}{c}{\underline{Faint: $R>21$}} \\
\colhead{} & \multicolumn{3}{c}{\underline{~~~~~All 4-band IRAC~~~~~}} & \multicolumn{3}{c}{\underline{~~~~~~~SEXSI X-ray~~~~~~~}} & \multicolumn{3}{c}{\underline{~~~~~All 4-band IRAC~~~~~}} & \multicolumn{3}{c}{\underline{~~~~~~~SEXSI X-ray~~~~~~~}}  \\
Sample & \colhead{Wedge} & \colhead{Total} & \colhead{\%} & \colhead{Wedge} & \colhead{Total} & \colhead{\%} & \colhead{Wedge} & \colhead{Total} & \colhead{\%} & \colhead{Wedge} & \colhead{Total} & \colhead{\%}}
\startdata
All     &   30 &  638 &  5\% $\pm$ 1\%   & 19 & 27  & 70\% $\pm$ 16\% & 374 & 1298 & 29\% $\pm$ 2\%  & 68  & 111 & 61\% $\pm$ 7\% \\
BLAGN   &      &      &                & 13 & 13 & 100\%  &     &     &                & 20 & 26 & 77\% \\ 
NLAGN   &      &      &                &  2 &  3 & 67\% &    &     &                &  7 & 11 & 64\% \\ 
ELG     &      &      &                &  1 & 6  & 17\% &     &     &                & 12 & 32 & 38\% \\
No ID   &      &      &                &  3 &  5 & 60\% &      &     &                & 29 & 42 & 69\% \\
\enddata
\tablecomments{The percentage errors represent the Poisson sampling uncertainties.}
\label{sexsiconclusion_tbl:wedgermag}
\end{deluxetable*}

\begin{deluxetable*}{lcccccccccc}
\tabletypesize{\footnotesize}
\tablecaption{Average properties of X-ray undetected IRAC sources.}
\tablecolumns{11}
\tablewidth{0pc}
\tablehead{Sample & $N$ & \multicolumn{3}{c}{\underline{~~~~~~0.5 -- 8 keV~~~~~~}} & \multicolumn{3}{c}{\underline{~~~~~~0.5 -- 2 keV~~~~~~}} & \multicolumn{3}{c}{\underline{~~~~~~2 -- 8 keV~~~~~~}} \\
 & & \colhead{Cts} & \colhead{Sig.} & $f_x$ & \colhead{Cts} & \colhead{Sig.} & $f_x$& \colhead{Cts} & \colhead{Sig.} & $f_x$ \\ 
 & & & & [$10^{-17}$] & &  &[$10^{-17}$]& & & [$10^{-17}$] }
\startdata
All & 1298  &  699.2 & 18.7 & 10.4 & 427.3 & 19.8 & 2.1  & 271.9 & 8.9 & 9.2 \\
Inside Wedge &  226  &  243.5 & 14.5 & 19.1 & 123.8 & 12.8 & 3.3  & 119.6 &  8.7 & 21.4 \\
Outside Wedge & 1072  &  455.7 & 13.6 & 8.4 & 303.4 & 15.7 & 1.9  &  152.3 &  5.6 & 6.4 \\
\hline
Inside Extended Wedge &  376  &  367.2 & 16.9 & 17.1 & 192.6 & 15.5 & 3.0  & 174.6 &  9.8 & 18.5 \\
Outside Extended Wedge & 922  &  332.0 &  10.9 & 7.2 & 234.7 & 13.3 & 1.7  &  97.3 &  3.9 & 4.8 \\
\enddata
\tablecomments{$N$ indicates the number of 
sources in the particular sample, Cts provides the number of 
net background-subtracted counts in each stacked sample, 
Sig. provides an estimate of the source significance ($S/\sqrt{B}$), 
and $f_x$ is the average X-ray flux
for the given band in units of $10^{-17} \fluxu$. See \S \ref{sexsiconclusion_sec:stackingmethod}.}
\label{sexsiconclusion_tbl:stackresults}
\end{deluxetable*}

\begin{deluxetable*}{cc}
\tablecaption{Wedge-selected stacked spectrum: implied \nh}
\tablecolumns{2}
\tablewidth{0pc}
\tablehead{\colhead{$z$} & \colhead{\nh} \\
  & \colhead{[$10^{22}$ cm$^{-2}$]}}
\startdata
0.0 & $1.3 \pm 0.2$ \\
0.5 & $3.5 \pm 0.6 $ \\
1.0 & $7.4 \pm 1.4 $ \\
1.5 & $14 \pm 3 $ \\
2.0 & $24 \pm 5 $ \\
3.0 & $48 \pm 9 $ \\
\enddata
\label{sexsiconclusion_tbl:stacknh}
\end{deluxetable*}


%
%
%
%
%

%
%

%
%

%
%
%

%
%

%
%
%

\end{document}